# iMaNGA: mock MaNGA galaxies based on IllustrisTNG and MaStar SSPs – I. Construction and analysis of the mock data cubes

Lorenza Nanni,[1]★ Daniel Thomas [ORCID],[1,2]★ James Trayford,[1]★ Claudia Maraston,[1] Justus Neumann [ORCID],[1] David R. Law,[3] Lewis Hill [ORCID],[1] Annalisa Pillepich [ORCID],[4] Renbin Yan [ORCID],[5] Yanping Chen[6] and Dan Lazarz[7]

[1]*Institute of Cosmology and Gravitation, University of Portsmouth, Dennis Sciama Building, Portsmouth PO1 3FX, UK*
[2]*School of Mathematics and Physics, University of Portsmouth, Lion Gate Building, Portsmouth PO1 3HF, UK*
[3]*Space Telescope Science Institute, 3700 San Martin Drive, Baltimore, MD 21218, USA*
[4]*Max-Planck-Institut fur Astronomie, Konigstuhl 17, D-69117 Heidelberg, Germany*
[5]*Department of Physics, The Chinese University of Hong Kong, Shatin, N.T., SAR 999077, Hong Kong SAR, China*
[6]*New York University Abu Dhabi, Abu Dhabi, PO Box 129188, UAE*
[7]*Department of Physics and Astronomy, University of Kentucky, 505 Rose St., Lexington, KY 40506-0057, USA*

## ABSTRACT

Galaxy formation and evolution simulations are essential tools to probe poorly known astrophysics processes, but particular care is needed to compare simulations with galaxy observations, as observed data need to be modelled as well. We present a method to generate mock galaxies from the hydro-dynamical IllustrisTNG simulations which are suited to compare with integral field spectroscopic observation of galaxies from the SDSS-IV/Mapping Nearby Galaxies at Apache Point Observatory (MaNGA) survey. First, we include the same instrumental effects and procedures as adopted in the acquisition and analysis of real data. Furthermore, we generate the galaxy spectra from the simulations using new stellar population models based on the MaNGA stellar library (MaStar). In this way, our mock data cubes have the same spatial sampling, cover the same wavelength range (3600–10 300 Å), and share the same spectral resolution ($R \approx 1800$) and flux calibration of real MaNGA galaxy spectra. In this first paper, we demonstrate the method over an early- and a late-type simulated galaxy from TNG50. We analyse the correspondent mock MaNGA-like data cubes with the same full spectral fitting code, FIREFLY, which was used for the observed spectra. We find that the intrinsic and recovered age and metallicity gradients are consistent within $1\sigma$, with residuals over all tassels consistent with 0 at the 68 per cent confidence level. We also perform the challenging test at comparing intrinsic and recovered star formation histories, finding a close resemblance between input and output. In follow-up papers, we will present a full simulated MaNGA-like catalogue ($\approx$10 000 galaxies) with a comprehensive comparison of TNG50 simulations to MaNGA observational results.

## 1 INTRODUCTION

Our understanding of galaxy formation and evolution, from a statistical point of view, has progressed considerably in the past few years thanks to both modern observational surveys and state-of-the-art hydro-dynamical simulations providing large statistical galaxy samples. From the observational side, we can now exploit data from modern spectroscopic surveys such as the Cosmic Assembly Near-infrared Deep Extragalactic Legacy Survey (CANDLES; Grogin et al. 2011), the Sloan Digital Sky Surveys (SDSS; York et al. 2000; Abazajian et al. 2003), and the Two-degree Field Galaxy Redshift Survey (2dFGRS; Baugh et al. 2004). Moreover, there has also been an increase in the quantity and quality of integral field spectroscopy (IFS) data set, with surveys such as Atlas3D (Cappellari et al. 2011), the Calar Alto Legacy Integral Field Area survey (CALIFA; Sánchez & et al. 2012), the Sydney-AAO Multi-object Integral field spectrograph (SAMI; Allen 2014), MUSE (Bacon et al. 2010), and Mapping Nearby Galaxies at Apache Point Observatory (MaNGA; Bundy et al. 2015). IFS surveys allow us to spatially resolve galaxies and study the local properties of their stellar populations and gas. Among the IFS surveys, MaNGA is the most extensive one, with a final catalogue of 10 000 unique galaxies in the local universe ($0.01 < z < 0.15$) in 2021.

On the theoretical side, large-scale hydrodynamical cosmological simulations of galaxies are now available, such as Illustris (Genel et al. 2014; Vogelsberger et al. 2014a; Sijacki et al. 2015), IllustrisTNG (Nelson et al. 2019a; see also Section 2.1), EAGLE (Schaye et al. 2014), and Horizon-AGN (Dubois et al. 2014), where baryonic matter and dark matter evolve together from high redshift ($z \approx 100$) to redshift zero, in large cosmological volumes. With large simulated samples, it is now possible to test galaxy formation and evolution theories and especially trying at constrain the parameters set to model poorly known processes.

★ E-mail: lorenza.nanni@port.ac.uk (LN); daniel.thomas@port.ac.uk (DT); james.trayford@port.ac.uk (JT)

A comprehensive approach to compare theoretical predictions against observational results is called 'forward modelling', where synthetic images and spectra are generated from the simulated data (e.g. Tonini et al. 2010; Snyder et al. 2015; Torrey et al. 2015; Trayford et al. 2015, 2017; Bottrell et al. 2017b; Guidi et al. 2018; Rodriguez-Gomez et al. 2018; Schulz et al. 2020).

In this paper, the first in a series, we present our method to construct realistic mock MaNGA-like IFU galaxy data cubes starting from the Illustris and IllustrisTNG simulations. However, our method is applicable to any hydro-dynamical galaxy simulations, such as EAGLE.

An original feature of our mocks is that galaxy spectra are obtained with stellar population models (and updates Maraston et al. 2020), based on the MaNGA stellar library (MaStar, Yan et al. 2019). In this way, we are able to construct mock galaxy spectra sharing the same spectral resolution, flux calibration and wavelength range as MaNGA observed galaxy spectra. Moreover, our data cubes have the same spatial sampling as MaNGA IFU. Other recent papers presenting mock MaNGA galaxies (Ibarra-Medel et al. 2018; Duckworth, Tojeiro & Kraljic 2019; Nevin et al. 2021) are based on standard population models instead of the MANGA-based ones we adopt here.

We also include MaNGA-specific observational effects, such as the effective Point Spread Function (ePSF; Law et al. 2016), and the typical noise of MaNGA observations. Finally, we follow the same steps as in the MaNGA Data Analysis Pipeline (DAP; Westfall et al. 2019) for re-binning data cubes and extracting kinematics and emission lines.

As a final step, we derive the stellar population properties of the mock data cubes via full spectral fitting of the mock spectra with the FIREFLY (Wilkinson et al. 2017) code, as is done for MaNGA galaxies (e.g. Goddard et al. 2016; Goddard et al. 2017; Lian et al. 2018; Oyarzún et al. 2019; Neumann et al. 2021, 2022).

In this paper, we demonstrate our method by analysing two illustrative mock galaxies that are typical of the MaNGA galaxy sample in terms of redshift and mass and compare their derived properties with the intrinsic ones. In future papers, we shall extend the approach to a MaNGA-like galaxy sample from TNG50 ($\approx$10 000 galaxies). In this way, we will be able to compare in the closest way possible what is predicted by *state-of-the-art* cosmological simulations, which are the outcome of our current theory on galaxy formation and evolution, to what we have observed thanks to the MaNGA survey.

This paper is organized as follows. Data and models used for this work are described in Section 2; our method for obtaining mock data cubes and their analysis is in Section 3. Results for the test-case mock galaxies are in Section 4 and conclusions follow in Section 5.

## 2 MODELS AND DATA

Here we provide a description of the models and observed data we use, namely the IllustrisTNG (Section 2.1) simulation data, the MaStar stellar population models (Section 2.3), and the MaNGA survey of galaxy spectra (Section 2.2);

### 2.1 The IllustrisTNG simulation suite

IllustrisTNG (Marinacci et al. 2018; Naiman et al. 2018; Nelson et al. 2018, 2019a; Springel et al. 2018; Pillepich et al. 2018b; Pillepich et al. 2019; Nelson et al. 2019b) is the successor of the Illustris project, a series of large-scale hydrodynamical simulations of galaxy formation (Genel et al. 2014; Vogelsberger et al. 2014a; Sijacki et al. 2015).

This new project retains the fundamental approach and physical models of Illustris, but expands the scope with simulations of larger volumes (up to 300 Mpc instead of 100 Mpc), at higher resolution (up to a mass resolution for the baryonic matter of $8.5 \times 10^4 M_\odot$ instead of $1.6 \times 10^6 M_\odot$), and with new physics. The principal new physics in IllustrisTNG is described in detail by Weinberger et al. (2017) and Pillepich et al. (2018a). It includes magnetic fields, and a dual-mode (i.e. thermal and kinetic) black hole (BH) feedback. Indeed, IllustrisTNG is a set of magneto-hydrodynamical (MHD) simulations where the magnetic field evolves through cosmic time from a minute primordial seed field at early times. The new kinetic BH feedback better regulates the stellar content of massive galaxies while preserving realistic halo gas fractions and alleviates the discrepancies identified in Illustris when compared to observations at the high end of the halo mass function ($10^{12}$–$10^{14} M_\odot$) (Pillepich et al. 2018a). The main physical processes these projects want to represent are the formation of dense gas clouds and stars; the evolution of stellar populations, their winds and energetic feedback; the explosions of supernovae; the formation of supermassive BHs and the macroscopic influence of their accretion, radiation and feedback; cooling, heating, radiative, and other microphysical processes in the turbulent, magnetized, multiphase interstellar medium; and the driving of galactic outflows of gas, energy, and heavy elements into the circum- and inter-galactic media. Indeed, these processes, acting across a wide range of spatial and time-scales, govern the formation and evolution of galaxies through cosmic time, regulating all their fundamental properties, such as their stellar content, star formation activity, gas content, heavy-element composition, morphological structure, and also the impact the galaxies could have on their surroundings. In particular, in these simulations, star formation occurs stochastically when the gas is characterized by a number density equal or above $0.13$ particle/cm$^{-3}$ following the Kennicutt–Schmidt law (Schmidt 1959; Kennicutt 1989) and assuming a Chabrier (2003) initial mass function (IMF).

In IllustrisTNG, three physical simulation box sizes are employed, characterized by cubic volumes of roughly 50, 100, and 300 Mpc side length (called TNG50, TNG100, and TNG300, respectively). They are also characterized by different resolution. For TNG50 (Pillepich et al. 2019; Nelson et al. 2019b), the gravitational softening and mass resolution for baryonic and dark matter are $\epsilon_{\rm gas,min} = 74$ and $\epsilon_{\rm DM,min} = 288$ pc, $m_{\rm bar} = 8.5 \times 10^4$ $M_\odot$, and $m_{\rm DM} = 4.5 \times 10^5$ $M_\odot$. During each run of the IllustrisTNG simulations, the Planck 2015 cosmological framework (Ade et al. 2016) is assumed (i.e. matter density parameter $\Omega_{\rm m} = 0.3089$; dark energy density parameter $\Omega_\Delta = 0.6911$; Hubble constant $H_0 = 100\,h\,{\rm km\,s^{-1}Mpc^{-1}}$, with $h = 0.6774$; amplitude of the matter power spectrum $\sigma_8 = 0.8159$; and spectral index $n_{\rm s} = 0.9667$) and 100 snapshots are saved from redshift 20.05 to redshift 0.0. In each snapshot of Illustris and IllustrisTNG, the Friends-of-Friends and SUBFIND algorithms identify haloes and subhaloes, respectively (for more details, see Springel et al. 2001; Nelson et al. 2015). For this work, we focus on subhaloes from TNG50 identified in snapshots from redshift 0.15 to 0.01, which roughly corresponds to the redshift range observed with MaNGA (see Section 2.2). We use TNG50 data because their high spatial resolution makes it possible to reproduce MaNGA-like data cubes that are characterized by a pixel size of 0.5 arcmin (so a spatial sampling going from $\approx$100 pc at $z \approx 0.01$ to $\approx$1.5 kpc at $z \approx 0.15$; Section 2.2). More details about the subhalo selection are given in Section 3.1 and a wider discussion will be included in our next paper, where the entire MaNGA-like catalogue is constructed from TNG50 simulation.

## 2.2 The MaNGA galaxy survey

MaNGA (Bundy et al. 2015) is the largest IFS survey of galaxies in the local volume to date, with a final sample of spatially resolved galaxy spectra for 10 010 unique galaxies at a median redshift of $z \sim 0.03$ (Abdurro'uf et al. 2022). MaNGA is part of the Sloan Digital Sky Survey (SDSS; Blanton et al. 2017), which been running since 2000 and more precisely of the SDSS-IV survey (Blanton et al. 2017), that concluded observations in 2020 August. The MaNGA spectra allow the analysis of stellar and gas properties over the 2D field of view (FoV). The MaNGA integral field spectrograph (IFS; Drory et al. 2015) was built around the SDSS 2.5-m telescope at Apache Point Observatory (Gunn et al. 2006) and the SDSS-BOSS spectrograph (Dawson et al. 2013; Smee et al. 2013), which has a wavelength range of 3600–10 300 Å and a spectral resolution $R \approx 1800$. The spectrograph is equipped with a red and a blue camera, with a dichroic splitting the light at roughly 6000 Å. The spectral resolution goes from $R = 1560$ at 3700 Å to 2270 at 6000 Å in the blue channel, while, in the red channel, it goes from $R = 1850$ at 6000 to 2650 at 9000 Å. We include these characteristics of the SDSS-BOSS spectrograph when we generate both the synthetic spectra (see Section 3.2.1) and the noise (see Section 3.4).

The MaNGA galaxy sample is divided into a 'Primary' and a 'Secondary' sample, following a 2:1 split. The 'Primary' sample observes a galaxy's optical extension out to 1.5 effective radius ($R_{\rm eff}$), whereas the 'Secondary' sample observes galaxies out to 2.5$R_{\rm eff}$ (Wake et al. 2017). In this work, we are interested in reconstructing data with the same quality as in the MaNGA primary sample (Section 3.4).

MaNGA is characterized by hexagonal-formatted fibre bundles, made from 2-arcsec-core-diameter fibres, employing dithered observations with Integral Field Units (IFUs) that vary in diameter from 12.5 (19 fibres) to 32.5 arcsec (127 fibres) (as described in table 2 in Bundy et al. 2015). Once the morphology is defined (see Section 3.4), the appropriate MaNGA IFU bundles are applied to our synthetic data cubes. MaNGA is characterized by a spatial resolution of 1.8 kpc at the median redshift of 0.037 (Law et al. 2016). To reach a target $r$-band signal-to-noise ratio (S/N) of 4-8 (Å$^{-1}$ per 2-arcsec fibre) at 23 AB mag arcsec$^{-2}$, which is typical for galaxy outskirts, MaNGA takes three sets of exposures, at dark time, around 15 minutes each, with 3 hours of integration time (Bundy et al. 2015; Yan et al. 2016). The typical fibre-convolved point spread function (PSF) has full width at half-maximum (FWHM) of 2.5 arcsec (Law et al. 2015). For each MaNGA data cube, the 'reconstructed' PSF, or effective PSF (ePSF), is also provided in different bands: this is an estimate of the spatial light profile of an unresolved point source during the MaNGA observation, depending on the exposure time and the observing condition at the considered band, including also dithering effects (Law et al. 2016). We use the ePSFs in the different bands when generating the mocks, as described in Section 3.4.

## 2.3 MaStar: SDSS-based stellar population models

Galaxy formation simulations employ stellar population models to generate the predicted spectra of model galaxies (Kauffmann, White & Guiderdoni 1993). The adopted stellar population model is a key ingredient in a galaxy formation simulation (Baugh 2006; Tonini et al. 2010) as it provides the link to the observables. At the same time, stellar population models are used to obtained the physical properties of data. It is then clear that the choice of model will play a central part in the comparison between galaxy simulations and real data.

In this work, we use stellar population models from Maraston et al. (2020) that adopt the MaNGA stellar library MaStar (Yan et al. 2019) for the description of stellar spectra as a function of effective temperature, gravity, and chemical composition in the population synthesis. MaStar is the new SDSS-based stellar library including 60 000 stellar spectra (Abdurro'uf et al. 2022),which was obtained using MaNGA fibre bundles and the BOSS optical spectrographs. This means that the stellar spectra and therefore the correspondent population models have the same wavelength range and resolution and flux calibration as the MaNGA galaxy spectra.

In this work, we use an updated version of the Maraston et al. (2020) models spanning a wider age range, in particular extending to young ages down to ∼3 Myr. The chemical composition of the models ranges between −2.25 and 0.35 dex in [$Z/H$], with 42 age values in the age grid, and 9 [$Z/H$] values for the metallicity grid. For each age and metallicity, there are also eight different assumptions regarding the IMF slope below 0.6 M$_\odot$, ranging between 0.3 and 3.8 in the notation in which the Salpeter's (1955) slope is 2.35. The synthetic spectra for the stellar particles in the simulated galaxies in TNG50 are constructed using these MaStar-based population models (see Section 3.2.1) assuming the Kroupa (2002) IMF. Since we assume a smoothly integrated (rather than stochastically sampled) IMF, stochastic effects are not accounted for. We note that the IllustrisTNG simulation itself does not account for stochasticity of the IMF neither in e.g. its feedback or enrichment processes. Therefore, the smoothly integrated treatment of the IMF is representative of the simulations. We also note that each spaxel contains typically several hundred spectra or more, which corresponds to a stellar mass of around 10$^6$ M$_\odot$. This makes any potential effects of stochasticity negligible (see Bruzual & Charlot 2003; Maraston 2005). The energetics and synthesis method are the same as in Maraston (2005) and Maraston & Strömbäck (2011).

Clearly, the strength of adopting these population models is in having the same spectral characteristics in the interpretative tools – the galaxy simulations – as well as in the observed data – the MaNGA galaxy sample. This is the highest degree of consistency that can be achieved when comparing models to data. By excluding offsets and effects due to the adoption of different spectral properties, the comparison between galaxy mocks and data will be revealing of the actual astrophysics process.

## 3 GENERATION OF MANGA MOCKS

In this section, we describe our method to generate mock-MaNGA galaxies from IllustrisTNG. We note that this method can be used with any other hydro-dynamical galaxy simulations with minimal changes. In the following, we briefly summarize the main steps of this process.

As a first step, we select galaxies in TNG50 snapshots, keeping track of their environment, as explained in Section 3.1. This is in view of a future analysis of stellar population properties as a function of the galaxy environment as in e.g. Goddard et al. (2017).

For each galaxy, we define a spherical aperture (Section 3.2) and calculate the light distribution produced by stellar populations and star-forming regions, if any (see Section 3.2.1). We also model dust effects, when dust is present (Section 3.2.2). As is well known, dust can significantly alter a galaxy SED, because of absorption, scattering, and re-emission processes. Radiative transfer (RT) calculations are needed to properly treat dust, but they are computationally expensive. Therefore we run an RT code post-processing when we know dust is present, to compute the extinction curve in each spaxel

**Table 1.** Main properties of the two example TNG50 galaxies 96-2 and 96-3.

|  | $M_i$ | $\log_{10} M_*$ | HMSR | Environment | $R_{\rm eff}$ | $n$ |
|---|---|---|---|---|---|---|
| 96-2 | −22.97 | 11.29 $M_\odot$ | 5.38 kpc | Mid-high | 7.13 kpc | 3.32 |
| 96-3 | −22.13 | 10.55 $M_\odot$ | 3.30 kpc | Mid-high | 4.86 kpc | 0.94 |

*Note.* From the left-hand side, the AB magnitude in the *i*-band $M_i$, the stellar mass $\log_{10} M_*$, and the HMSR as given by the TNG50 data, followed by the environment (Section 3.1), the effective radius $R_{\rm eff}$, and the Sérsic index $n$ (Section 3.3), as determined in this work.

(see Section 3.2.2). Otherwise, we proceed with our own code to construct synthetic MaNGA data cubes (see Section 3.2.1).

When the data cube is ready, we construct a synthetic SDSS r-band image, in order to probe the galaxy morphology, as done for MaNGA galaxies (Blanton 2006). Then we employ the MaNGA fibre bundle configuration containing up to $1.5 R_{\rm eff}$ of the galaxy, see Section 3.3.

We further post-process the spectra within the selected FoV by adding the effects of the ePSF in the different bands, and the noise for each spaxel based on MaNGA data from the Data Reduction Pipeline (DRP; Law et al. 2016, 2021); see Section 3.4.

As a final step, we proceed as we would for an observed-MaNGA galaxy, following the steps in the MaNGA DAP, namely the adaptive binning scheme implemented by the Voronoi algorithm with target $S/N_g > 10$ is applied (Cappellari & Copin 2003; Westfall et al. 2019) and the code PPXF is run to obtain the kinematics and model the emission lines (Cappellari 2017; Belfiore et al. 2019; Westfall et al. 2019), as discussed in Section 3.5.

To illustrate the pipeline, we focus on two example galaxies from TNG50, with stellar masses ranging from 10.5 to 11.3 in $\log M/M_\odot$ and $z \approx 0.03$ – as typical of MaNGA observations (see Table 1). These objects have ID = 2 and 3 and belong to snapshot 96 in TNG50 ($z \approx 0.03$), henceforth they will be referred to as galaxies 96-2 and 96-3.

### 3.1 Galaxy selection and galaxy environment

As explained in Section 2.1, the SUBFIND algorithm determines the galaxies present in the simulated cosmological volumes in each saved snapshot of Illustris and IllustrisTNG simulations. To construct MaNGA-like galaxies from TNG50 simulations, we select galaxies in the MaNGA redshift range (see Section 2.2), i.e. between $z \approx 0.15$ and $\approx 0.01$, which corresponds to snapshots 98 and 88. Among these, we further select those galaxies with an absolute magnitude (AB) in the *i* band between −23 and −17, as for the MaNGA target (Section 2.2), and with more than 10 000 stellar particles, to ensure they are sufficiently resolved to reconstruct dust effects (e.g. as done in Schulz et al. 2020). A more extended discussion of our selection criteria for the whole TNG50 will be part of our next paper.

For the selected TNG50 galaxies, we also examine their local environment as follows. We estimate the local galaxy (projected) overdensity using the distance to the *N*th nearest neighbour, $d_N$. This distance has often been used as a measure of the overdensity, with *N* typically varying from 3 to 10 (Muldrew et al. 2011; Etherington & Thomas 2015). We define the environment of a galaxy in terms of the dimensionless overdensity, $1 + \delta$, as

$$1 + \delta = 1 + \frac{\Sigma_N - \langle\Sigma\rangle}{\langle\Sigma\rangle}, \quad (1)$$

where $\Sigma_N$ is the surface number density described using *N*th neighbour method ($\Sigma_N = N/\pi d_N^2$), and $\langle\Sigma\rangle$ is the mean surface density of galaxies within the snapshot, projecting the subhaloes in the cosmological volume along the *z*-axis. The galaxy environment is given by $\log(1 + \delta)$. Following Goddard et al. (2017), who looked at the environment for MaNGA galaxies using SDSS images, we select $N = 5$ and use $10^7 M_\odot$ as a threshold for stellar mass, since it is roughly the stellar mass limit in SDSS, to compute the surface number density and the surface density of galaxies within the snapshot. As in Goddard et al. (2017), we split the galaxy environment distribution into quartiles to define four different environmental densities and assign our galaxies to these groups as follows:

(i) **low-density environment**: $\delta <$ 25th percentile;
(ii) **mid-low-density environment**: 25th percentile $< \delta <$ 50th percentile;
(iii) **mid-high-density environment**: 50th percentile $< \delta <$ 75th percentile;
(iv) **high-density environment**: $\delta >$ 75th percentile.

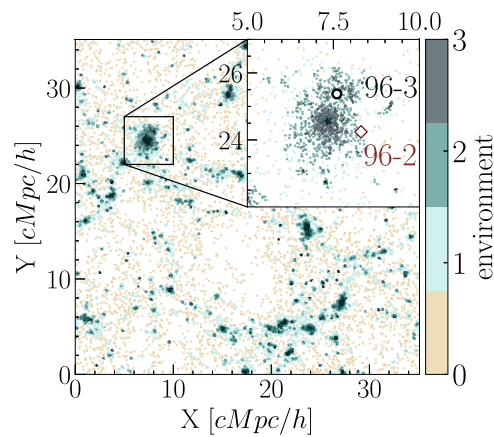

**Figure 1.** The 2D galaxy distribution in TNG50 at $z \approx 0.03$ (i.e. snapshot 96), projecting along the *z*-axis of the comoving cosmological volume. The dots represent the position of the centre of mass of every galaxy in snapshot 96 with a stellar mass above $10^7 M_\odot$. The dots are also coloured according to the density of their environment, defined with the fifth neighbour method. The positions of the centre of mass of the galaxies with ID = 2 and 3 at $z \approx 0.03$ (referred as 96-2 and 96-3 in figure, respectively) are reported in the zoom-in box in the upper right-hand corner (the empty red diamond and the empty blue circle, respectively): both are in a mid-high-density environment.

In summary, the galaxies selected are characterized by snapshot, id, redshift, environment (as 1, 2, 3, or 4), half-mass stellar radius (HMSR), total stellar mass, and $M_i$ (see Table 1).

Fig. 1 shows the 2D galaxy spatial distribution in the snapshot 96 of TNG50 ($z \approx 0.03$), projecting along the *z*-axis of the comoving cosmological volume, and colour-coded according to the environment, computed as just discussed. In particular, the figure reports the position of the centre of mass of the DM subhaloes, for all subhaloes with $M^* > 10^7 M_\odot$. A zoom-in view around the two selected subhaloes, which turned out to reside in a mid-high-density environment, is given in the upper right-hand corner.

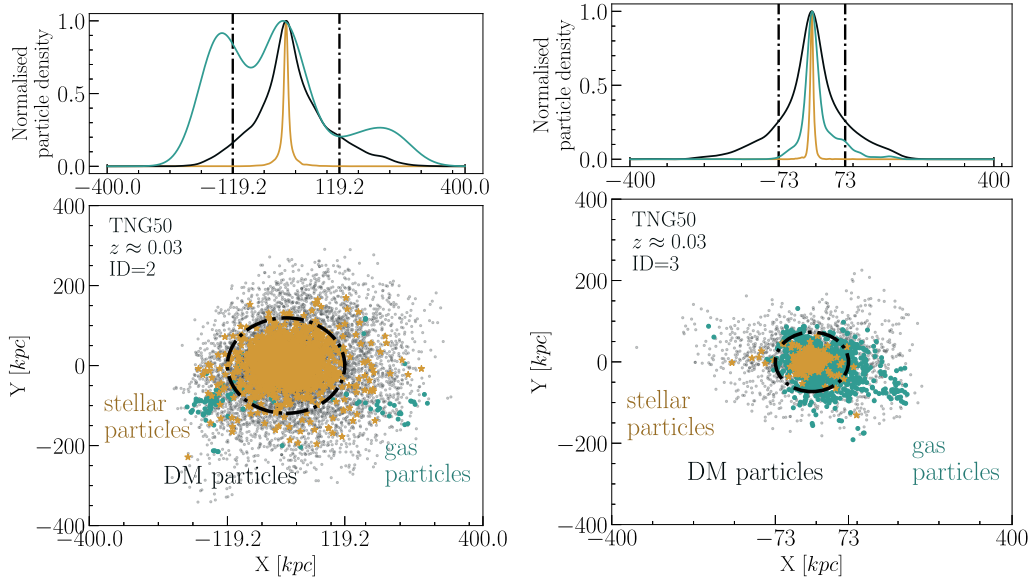

**Figure 2.** The spatial distributions of DM, gas and stars in two example galaxies in TNG50 with ID = 2 and 3 in snapshot 96, so $z \approx 0.03$ (henceforth they will be referred to as galaxies 96-2 and 96-3 in following captions) Specifically, the DM, gas, and stars 2D spatial distribution is reported (black dots, blue circles, and yellow stars, respectively), projecting along the $z$-axis, centred in the galaxy centre-of-mass frame. For more readable plots, 1/500 DM particles and 1/100 gas cells and stellar particles are reported. The dash-dotted black circles identify the projection of the spherical apertures having a radius equal to 15 times the HMSR. On top, the DM, gas, and stellar normalized 1D spatial density distribution (black, blue, and yellow solid line, respectively), along the comoving $x$-axis, centred in the centre-of-mass frame. The vertical dash–dotted lines illustrate where the defined spherical apertures extract the particles (15 times the HMSR; the HMSR values are reported in Table 1).

### 3.2 Data cube creation

To post-process and analyse simulated galaxies, a spherical aperture is used to define the galaxy properties. In preview analysis, different apertures have been used. For example, in Nelson et al. (2018) and Vogelsberger et al. (2020), a fixed spherical aperture of 30 kpc is used to extract the galaxy particles and their information. In Schulz et al. (2020), all the particles within twice the half-mass stellar radius are extracted to define the galaxy properties. In Rodriguez-Gomez et al. (2018), the image for each galaxy is created looking within a spherical aperture of 15 times the half-mass stellar radius. In this work, we define a galaxy by all gravitationally bound stellar particles and gas cell – so all SUBFIND resolution elements – found within a spherical aperture of 15 times the HMSR (see Table 1), following Rodriguez-Gomez et al. (2018).

Fig. 2 shows the normalized 1D particle spatial density (upper panels) and 2D particle spatial distribution (bottom panels), for the selected galaxies, the apertures used to extract the particles and their properties. In particular, the bottom panels illustrate the 2D spatial distribution of DM, gas, and stellar particles in galaxy 96-2 (left-hand panel) and galaxy 96-3 (right-hand panel), as given by the simulations, namely all particles the SUBFIND algorithm identifies, projecting along the $z$-axis centred in the subhalo centre-of-mass frame. The dash–dotted circles represent the projection of the spherical apertures we use to extract the particles, so only particles within the dash–dotted lines define galaxies 96-2 and 96-3 in this work. The upper panels, instead, report the normalized 1D spatial density, along the comoving $x$-axis, for DM, gas, and stellar particles, and the vertical dash–dotted lines show where the spherical apertures extract the particles.

Once we extract the stellar and gas cells within the spherical apertures, we build the galaxy light.

#### 3.2.1 IMASTAR pipeline

IMASTAR is a code we developed to generate synthetic data cubes mimicking an IFU instrument. In IMASTAR, we mimic a MaNGA-type integral field unit (IFU) bundle, with pixel size of 0.5 arcsec and an FoV of $150 \times 150$ arcsec$^2$, so 300 spaxels per side. We position our virtual IFU at the galaxy redshift along the $z$-axis, in the centre-of-mass frame. Observing the simulated galaxies along the $z$-axis means having a random viewing angle for each of them, as they have a random spatial distribution and orientation in the cosmological volume of the simulation.

Each particle in the spaxel is then associated with a spectrum, according to its age, metallicity, and stellar mass as given by IllustrisTNG, assuming different models according to the particle age as described below.

For particles older than 4 Myr, we use MaStar stellar population models (see Section 2.3). Being based on the MaStar models, our synthetic spectra have the same resolution, wavelength range, and flux calibration of MaNGA observations (see Sections 2.2–2.3). Younger stellar particles are treated as star-bursting regions, modelling their SEDs with the MAPPINGSIII photoionization code (MIII; Groves et al. 2008). In works such as Trayford et al. (2017), Rodriguez-Gomez et al. (2018), and Schulz et al. (2020), all stellar particles younger than 10 Myr are treated as starbursting regions exploiting MIII, in order to include the emission lines from the surrounding photo-dissociation regions (PDRs), as well as absorption by gas and dust in the birth cloud. As MaStar models are available down to 3.2 Myr, we adopt them between 4 and 10 Myr, since the emission lines from the surrounding PDRs become negligible at these ages. To model spectra from MIII, we have to set the Star Formation Rate (SFR), the metallicity, the compactness parameter $C$, the interstellar medium (ISM) pressure $P_0$, and the PDR covering

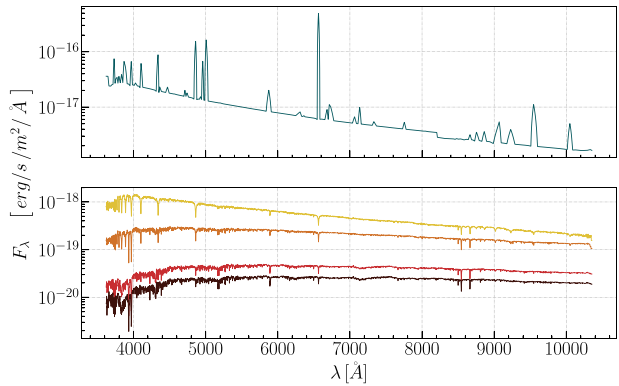

**Figure 3.** Upper panel: MIII spectrum for a star-forming region with SFR = $0.1 M_\odot/\text{yr}$, $[Z/H] = 0.$, $\log C = 5$, $\log[(P_0/k_B)/\text{cm}^{-3}\text{K}] = 5$, and $f_{\text{PDR}} = 0.1$ at $z \approx 0.03$. This modelling that mimics gas emission lines from star-forming regions, is applied to simulated stellar particles younger than 4 Myr. Lower panel: MaStar spectrum for stellar particles with initial mass $10^6 M_\odot$, $[Z/H] = 0.$, and ages $t = 100$ Myr, 500 Myr, 5 Gyr, and 10 Gyr (in yellow, orange, red, and brown, respectively), at redshift $z \approx 0.03$. This purely stellar population modelling is applied to stellar particles older than 4 Myr.

fraction $f_{\text{PDR}}$. To set these parameters, we follow Trayford et al. (2017), as

(i) the SFR is given by the stellar particle initial mass divided by 10 Myr, assuming that the SFR is constant over 10 Myr;

(ii) the metallicity is fixed at the value given by the simulation for that stellar particle;

(iii) $\log C = 5$ and $\log[(P_0/k_B)/\text{cm}^{-3}\text{K}] = 5$ as in Groves et al. (2008);

(iv) $f_{\text{PDR}} = 0.1$, as done by Camps et al. (2016), who was able to reproduce the far-infrared properties of observed nearby galaxies using the EAGLE simulations.

Fig. 3 shows example spectra from IMASTAR based on MaStar and MIII stellar population models. The upper panel reports the spectrum of a star-forming region with a stellar particle younger than 4 Myr (with SFR= $0.1 M_\odot \text{yr}^{-1}$ and $[Z/H] = 0$, and the other parameters as just discussed). Strong emission lines can be appreciated. The bottom panel shows spectra for stellar particles with ages $t = 100$ Myr, 500 Myr, 5 Gyr, and 10 Gyr for an initial mass of $10^6 M_\odot$ and $[Z/H] = 0$. These are MaStar-based absorption-line spectra.

The total synthetic spectrum for each spaxel is the sum of all spectra within that spaxel. In case of central spaxels, this can total to up to $\sim 10\,000$ spectra.

For each spaxel, we recover also the kinematics from TNG50 to include the kinematic effects in the spectra. To add the kinematic effects to each spaxel, we follow two different approaches, depending whether star-forming regions are present or not. When we only have stellar particles older than 4 Myr, we only reconstruct the stellar kinematics. Considering both the peculiar velocity along the line of sight (LOS, the $z$-axis in our construction) and the galaxy redshift, the spectrum in each spaxel is shifted by the Doppler effect. Additionally, the spectrum in each spaxel is downgraded according to the stellar velocity dispersion and the resolution of the MaNGA instrument exploiting a 1D Gaussian kernel with an FWHM defined as

$$\text{FWHM} = \lambda \times \frac{\sqrt{\sigma_{v_z}^2 + \sigma_{\text{inst}}^2}}{c}, \quad (2)$$

where $\sigma_{v_z}$ is the stellar velocity dispersion, computed as the standard deviation of the stellar peculiar velocity along the LOS in the spaxel, $\sigma_{\text{inst}}$ is equal to $c/R_{\text{inst}}/2.355$, with $R_{\text{inst}}$ being the wavelength-dependent SDSS-BOSS spectrograph resolution computed as the median resolution of a selection of stellar spectra from MaStar (see section 5.7 in Maraston et al. 2020), and $c$ is the speed of light.

When there are stellar particles younger than 4 Myr, as for galaxy 96-3, we generate the rest-frame zero-velocity synthetic IFU data cubes for the star-forming regions and the older stellar component separately. Then, their kinematics are recovered independently: for stellar particles older than 4 Myr, spectra are shifted and downgraded as just discussed; for stellar particles younger than 4 Myr, we additionally reconstruct the kinematics of simulated gas cells with SFR>0 found within the same spaxel. Indeed, the kinematics of star-forming regions resemble the cold gas kinematics. The spectrum is then Doppler shifted according to the mean peculiar velocity along the LOS from both stars and gas. The spectrum is also downgraded by both the instrumental resolution and velocity dispersion, computed from the peculiar velocity along the LOS of both stellar particles and gas if any, following again equation (2). The emission lines are already broadened for their typical thermal velocity in the MIII models (Groves et al. 2008). Finally, the two synthetic data cubes are summed together, spaxel by spaxel.

Fig. 4 illustrates, for both example galaxies, slices of the final total synthetic data cubes (left-hand panels) and stellar kinematics along the LOS, namely stellar velocity (central panels) and stellar velocity dispersion (right-hand panels). The upper row of Fig. 4 refers to galaxy 96-2, which is composed by only stellar particles older than 4 Myr, while the lower row refers to galaxy 96-3, containing stellar particles younger than 4 Myr. From the left-hand panels, we see that galaxy 96-2 has a higher flux and is more diffuse than galaxy 96-3. Galaxy 96-2 exhibits a hot-kinematical structure overall, with a higher stellar velocity dispersion than galaxy 96-3, which instead shows spiral-like structure and a rotationally supported, cold-kinematical structure and a low velocity dispersion.

Since galaxy 96-3 includes stellar particles younger than 4 Myr, we reconstruct stellar and gas kinematics as just discussed. Fig. 5 illustrates the velocity and velocity dispersion maps for all gas cells within the spherical aperture (upper panels) and for the gas cells with non-null SFR (lower panels).

Fig. 6 shows the final total spectrum in a spaxel for galaxy 96-3 (upper panel), and the effects of the kinematics (bottom panel). Specifically, the spectrum in the upper panel is the sum of the MIII and MaStar spectra in that spaxel after the Doppler shifting and the downgrading, as explained above. The bottom panel is a zoom-in of the spectrum between 4000 and 4500 Å, where the MIII and MaStar spectra are shown before and after the Doppler shift and the downgrading effects are included, to demonstrate the effects of kinematics on spectra: the absorption and emission lines appear shifted and broadened.

### 3.2.2 Adding dust effects with SKIRT

Dust can significantly alter the SED of a galaxy. The dust in the interstellar medium (ISM) is not tracked in IllustrisTNG simulations, but it is reasonable to assume that the metal distribution within gas cells traces the ISM dust distribution (as done, for example, in Trayford et al. 2017; Rodriguez-Gomez et al. 2018; Schulz et al. 2020). In particular, dust traces the cold metal-rich gas, which is associated with star formation. Therefore, dust is only associated with those gas cells with either a temperature below $T = 8000$ K or a non-zero star formation (as in e.g. Camps et al. 2016; Trayford et al. 2017). As dust extinction is already included within the star-forming regions in the MIII models, we remove dust attenuation of gas cells within 100 pc around MIII particles in the simulation, to avoid double counting the dust associated with stellar birth clouds.

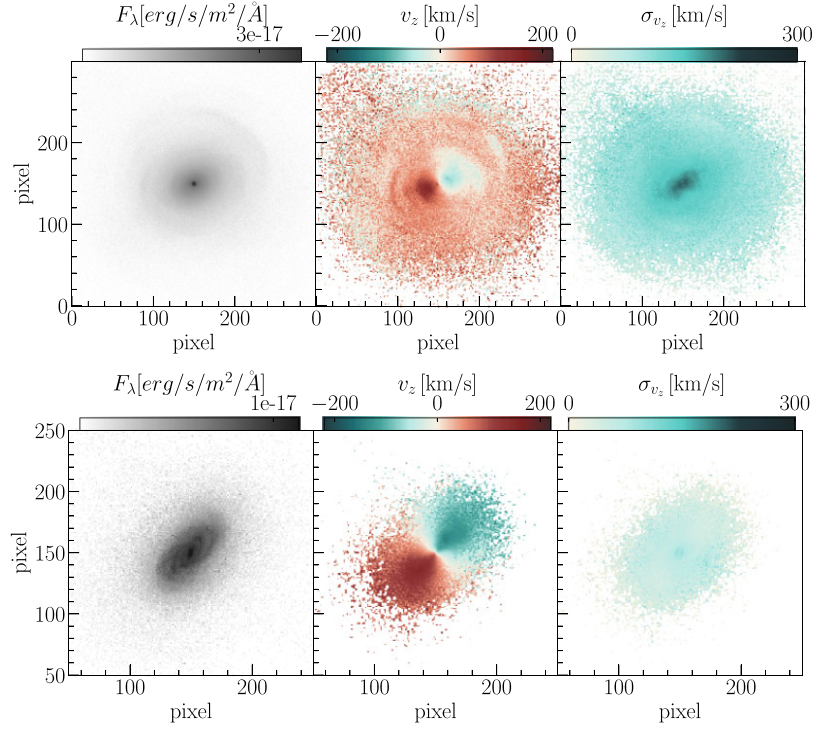

**Figure 4.** Slice of the synthetic data cubes at $\lambda = 6322$ Å from the IMASTAR code and stellar kinematics for the two example galaxies 96-2 (upper panels) and 96-3 (bottom panels). Stellar velocity maps are computed from the mean stellar peculiar velocity component along the LOS in each spaxel. Stellar velocity dispersion maps are the standard deviation of the peculiar stellar velocity along the LOS in each spaxel.

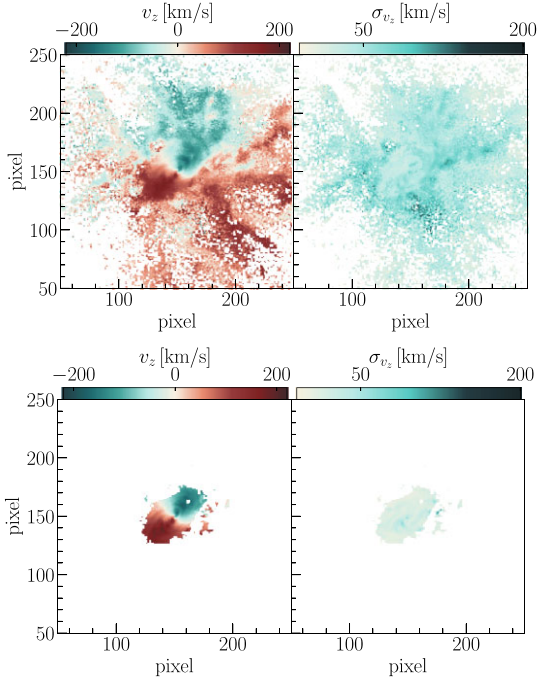

**Figure 5.** Gas kinematics for the example galaxy 96-3 at $z \approx 0.03$. Upper panels show the velocity (left-hand panel) and velocity dispersion (right-hand panel) maps for all gas particles, while the lower panels focus on gas cells with SF > 0.

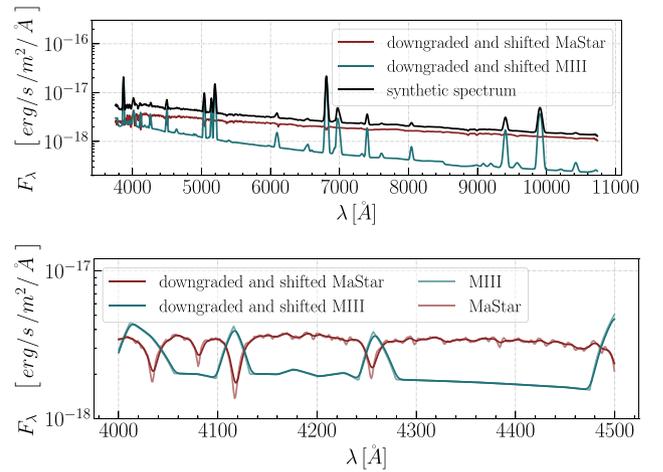

**Figure 6.** The final total synthetic spectrum and a zoom-in between 4000 and 4500. Å, to show the effects of the shifting and downgrading over the zero-velocity synthetic spectra, for the example galaxy 96-3. In the *upper panel*, the black solid line is the final synthetic spectrum that is the sum of stellar components older than 4 Myr obtained with MaStar SSPs models (red solid line), and those younger than 4 Myr based on MIII SSPs plus photo-ionization models (blue solid line). Both MaStar and MIII spectra are shifted according to the correspondent reconstructed kinematics in that spaxel. Spectra are also downgraded assuming both the instrumental resolution and the velocity dispersion along the LOS in that spaxel. For MIII, the emission lines are also broadened for their typical thermal velocity in the MIII models. The bottom panel shows the effect of shifting and downgrading over the zero-velocity synthetic spectra.

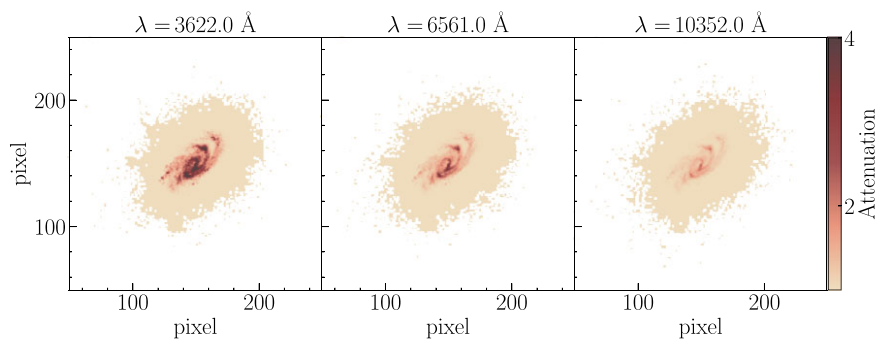

**Figure 7.** Attenuation maps at three different wavelengths: 3622, 6561, 10352 Å, for galaxy 96-3, as reconstructed from SKIRT outputs in the *FullInstrument* option. Maps are defined as the ratio between the slices of the data cube for the system with dust component and without, at the same wavelength.

SKIRT[1] (Baes et al. 2011; Baes & Camps 2015) is a public 3D Monte Carlo radiative transfer code for simulating the effect of dust on radiation from astrophysical systems. In this work, we use SKIRT to perform dust radiative transfer simulations for the galaxies in our sample that contain dust as explained above. In particular, we include MaStar stellar population models as a novelty into the SKIRT library. Also in SKIRT, we use MIII to describe the emission from stellar particles younger than 4 Myr.

Following Trayford et al. (2017), we choose the multicomponent dust mix of Zubko, Dwek & Arendt (2004) including graphite grains, silicate grains, and polycyclic aromatic hydrocarbons (PAHs). The intrinsic extinction properties of dust grains are assumed to follow Zubko et al. (2004). We fix the dust-to-metal mass ratio as

$$f_{\rm dust} = \frac{\rho_{\rm dust}}{Z \rho_{\rm gas}} = 0.3, \quad (3)$$

where $Z$ is the metallicity of the gas particle, and $\rho_{\rm dust}$ and $\rho_{\rm gas}$ are the dust and gas density, respectively. Equation (3) is motivated by the finding that a constant $f_{\rm dust}$ has been observed in many environments (e.g. Mattsson et al. 2014), with the value 0.3 derived from observations (e.g Draine et al. 2007). We can then set the dust mass from the mass of the gas cells in the simulation as $m_{\rm dust} = 0.3 Z m_{\rm gas}$.

Following what has been done in Schulz et al. (2020), we perform the radiative transfer calculation directly on a three-dimensional Voronoi mesh. Indeed, the IllustrisTNG simulation suite is based on the AREPO code (Springel 2010). In this way, we can directly reconstruct the gas distributions exactly as they are implemented in the hydrodynamic solver in order to simulate the dust effects in the system. Indeed, with this approach, the coordinates of the gas cells are the mesh-generating points, which are used by SKIRT to reconstruct the Voronoi mesh inside this volume using the VORO++ open source library (Rycroft 2009) for computing Voronoi tasselations.

Once the light sources and the dust distribution are defined, SKIRT simulates dust absorption, scattering, heating, and re-emission processes. SKIRT also includes a suite of tools mimicking astronomical instrumentation, such as imaging and 3D spectroscopic devices. Here we mimic a MaNGA-type IFU, using the same approach to generate the mocks (as discussed in Section 3.2.1).

We run SKIRT for each galaxy using the *FullInstrument* option, namely we output one data cube with only stellar particle emission and one with also dust included. In this way, we construct a coarse absorption curve in each spaxel. We then apply these attenuation curves to the IMASTAR code output by interpolating them in wavelength with a cubic spline interpolation at the considerably finer resolution of the MaStar model spectra. In the assumption that attenuation curves vary slowly and smoothly relative to our chosen spectral resolution, we are able to model dust effects.

With this method, we construct the attenuation curve for each spaxel, so that it is geometry-dependent. Indeed, our attenuation curves are constructed reproducing the 3D geometry of dust and stars in SKIRT. We do reconstruct coarse attenuation curves, hence we assume dust is smoothly distributed. This means that we do not reconstruct the fine details in wavelengths of the attenuation curves. A smooth attenuation curve is also what is assumed in the actual recovery of stellar populations using FIREFLY (see Section 4.1) as with most approaches for recovering intrinsic stellar spectra. Ideally, we would run the RT code at the MaNGA wavelength resolution. Instead, we proceed in this way mainly because any RT simulation is characterized by stochastic processes that lead to Poisson noise. Using a high number of photon packets would make the Poisson noise negligible only in the highly dense regions. Instead, with our method, we are able to reproduce accurately the noise characteristic of MaNGA, as explained in Section 3.4. Moreover, it is computationally unfeasible to run SKIRT with the MaNGA wavelength grid at such relatively high resolution.

Fig. 7 shows the reconstructed attenuation maps at different wavelengths ($\lambda$ =3622, 6561, and 10 352 Å, as an example) for galaxy 96-3, defined as the ratio between the data cube slice at the given wavelength as produced by SKIRT for the system without and with dust. As expected, the attenuation increases towards shorter wavelengths. When the attenuation is equal to 1, the dust does not produce any effect on the light from stellar particles. We can also notice how the central part of the galaxy is most affected by dust.

The top panel of Fig. 8 displays the attenuation curve in a spaxel, given as the cubic spline interpolation of SKIRT outputs at eight different wavelengths. The best fit for the Calzetti et al. (1994) extinction curve is also shown, which corresponds to $E(B - V) = 0.24$ as the best-fitting value. The bottom panel demonstrates how the spectrum in that spaxel is modified by applying the reconstructed attenuation curve.

Since we use cold gas ($T < 8000$ K) or gas with non-null SF to trace the presence of dust, if gas cells with these properties are absent as for galaxy 96-2, the SKIRT pipeline is not run.

---

[1] SKIRT documentation: http://www.skirt.ugent.be.

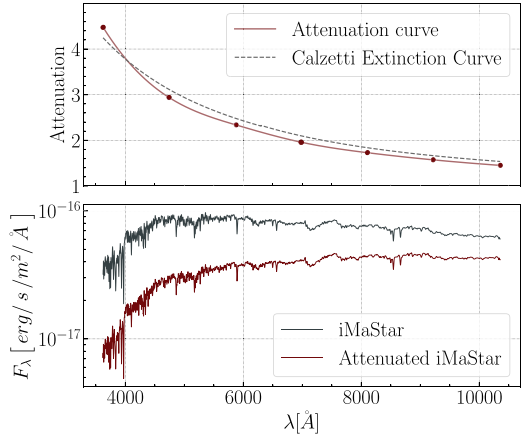

**Figure 8.** Attenuation curve (top panel) and its effect (lower panel) over a spaxel from the synthetic data cube generated with IMASTAR (Section 3.2.1) for galaxy 96-3. In the top panel, the red dots are the ratio between the flux from the system without and with dust, outputted by SKIRT in the *FullInstrument* mode. The red solid line is the cubic spline interpolation of the attenuation values represented by the red dot points. Fitting the red dots with a Calzetti, Kinney & Storchi-Bergmann (1994) law, we find $E(B-V) = 0.24$ as the best fit. The dotted grey line is the Calzetti et al. (1994) extinction curve for this value. In the bottom panel, the original spectrum from the IMASTAR code (black line) (Section 3.2.1) and the same one once attenuation is applied (red line).

### 3.3 Optical morphology and FoV selection

Once we have the synthetic data cubes, we quantify their morphology with STATMORPH (Rodriguez-Gomez et al. 2018), a Sérsic (Sérsic 1963; Sersic 1968) 2D fitting code. Morphologies allow us to determine the effective radius $R_{\rm eff}$, which is crucial for producing MaNGA mock observations as the MaNGA Primary sample refers to spectra within $1.5 R_{\rm eff}$ (see Section 2.2).

To study the morphology, we first create a galaxy synthetic image in the *r* band, which is the band at which galaxy $R_{\rm eff}$, position angles, and ellipticities are measured (Blanton 2006). To this end, we include the *r*-band SDSS filter response, the SDSS PSF, and noise, as in Bottrell et al. (2017a), where SDSS images from simulated galaxies in Illustris100 (Genel et al. 2014; Vogelsberger et al. 2014b; Nelson et al. 2015; Sijacki et al. 2015a) are produced. The first step is to convolve our synthetic data cubes with the SDSS *r*-band hotometric filter response (Gunn et al. 1998). We construct the SDSS PSF function as the sum of two Gaussian profiles with an FWHM equal to 1.4 and 2.8 arcsec, respectively (following Law et al. 2015). We add a random noise to each pixel in order to have a minimum S/N of 20, so that the source can be easily identified and its morphology determined. We note that the proper construction of MaNGA-like galaxies with a more realistic modelling of PSF and noise is done over the entire synthetic data cubes, as will be discussed in Section 3.4.

STATMORPH requires the image, the segmentation map, and the weighted map. We create the segmentation map with PHOTUTILS (Bradley et al. 2020), that identifies a source from the background noise, applying a convolution filter to separate low-surface-brightness objects from spurious detections. In particular, we use a detection threshold of $1.5\sigma$ above the local sky background and a minimum group number of 8 pixels to trigger a detection. The weighted map, or sigma map, is simply computed as the standard deviation of each pixel value. With these inputs, the Sérsic 2D fitting code produces as output the best-fitting values for the morphological properties of the galaxy. Fig. 9 presents the *r*-band image, Sérsic 2D best-fitting model, and the residual, for both galaxies 96-2 (upper panels) and 96-3 (bottom panels). As expected, we have higher residuals where structures such as spiral arms are present. Galaxy 96-2 has a Sérsic index $n = 3.26$ consistent with being an early-type galaxy (ETG), while galaxy 96-3 has a Sérsic index $n = 0.94$, consistent with a late-type galaxy (LTG) morphology; both these values are reported in the central panels, together with the $R_{\rm eff}$ (∼22pixels, so around 8 kpc, for galaxy 96-2, and ∼14 pixels, so around 5 kpc for galaxy 96-3), the amplitude, the centre of the Sérsic 2D profile, and the half-light ellipse. These values are also reported in Table 1.

In Table 1, we report key properties of the selected galaxies. From the simulations, we know their magnitude in the *i* band, their stellar mass and their HMSR radius, while, from our analysis so far, we also know their environment, their effective radius and their Sérsic index.

### 3.4 Mock MaNGA data cubes

Once we know the sky size of the $1.5R_{\rm eff}$, we can select the MaNGA FoV (see Section 2.2) that would be used to observe a galaxy with that sky size. For our example galaxies, FoV = 32.5 and 22.5 arcsec for 96-2 and 96-3, respectively. We then apply this FoV bundle to our produced synthetic data cubes Section 3.2. At this stage, we add realism to our synthetic data cubes, produced as discussed in Section 3.2, by modelling the noise and convolving for the ePSF in the different wavelength bands.

We generate the noise in each spaxel as a Gaussian perturbation, defined as

$$\frac{{\rm d}F_{{\rm spaxel},\lambda}}{F_{{\rm spaxel},\lambda}} = \frac{\sqrt{F_{1.5,\lambda}}}{{\rm S/N}_{1.5}(\lambda) \times \sqrt{F_{{\rm spaxel},\lambda}}}, \quad (4)$$

where $F_{{\rm spaxel},\lambda}$ is the flux in the selected spaxel and at given wavelength $\lambda$, $F_{1.5,\lambda}$ is the flux at $1.5R_{\rm eff}$ at given wavelength $\lambda$, and ${\rm S/N}_{1.5}(\lambda)$ is the S/N as a function of $\lambda$ at $1.5R_{\rm eff}$, computed from MaNGA observations. To compute the ${\rm S/N}_{1.5}(\lambda)$ in MaNGA observations, we select 100 random observed galaxies in the MaNGA primary sample, and measure the S/N as a function of wavelength for all spaxels at ∼$1.5R_{\rm eff}$ for the *LOGCUBE* produced by the DRP (Law et al. 2016).

Fig. 10 reports the averaged S/N for all spaxels at ∼$1.5R_{\rm eff}$ for 100 MaNGA *LOGCUBE* as a function of $\lambda$, after the application of *DRP3PIXMASK* mask to reject bad spaxels (Law et al. 2016). As expected from the SDSS-BOSS spectrograph, the S/N is lower at the edge of the wavelength range. The effect of wavelength splitting is visible at roughly 6000 Å (see Section 2.2).

Fig. 11 illustrates how spaxel spectra are modified by the noise for the example galaxy 96-3. The noise-free spectra in the central spaxels (top panel) and ∼$1.5R_{\rm eff}$ (bottom panel) in the synthetic data cube and after noise is included are shown. It is possible to appreciate that we obtain larger noise at the edges of the wavelength range, consistent with observations with the BOSS spectrograph (see discussion in Section 2.2). The noise also increases towards the outskirt of the fibre bundle configuration, as for MaNGA observations (see Section 2.2). The use of this method for modelling the noise, in combination with MaStar population models, makes the resolution and S/N of our simulated galaxies as close as possible to real MaNGA observations.

In the MaNGA DRP outputs, each data cube has four extensions illustrating the reconstructed PSF in the *griz* bands (Law et al. 2016). Since the requirements on exposure time and the observing conditions in MaNGA are designed to minimize ePSF variations between observations, the ePSF in each band does not vary a lot among different data cubes. Therefore, we construct representative ePSFs for each band, by averaging 100 randomly selected MaNGA data cubes from the DRP output. Fig. 12 shows the reconstructed average *r*-band ePSF. The bottom panel illustrates the 2D kernel we use for the *r* band, while the top panel reports its projection along the slice highlighted with a grey band, as an example of its 1D profile. The 2D reconstructed kernel is normalized to 1.

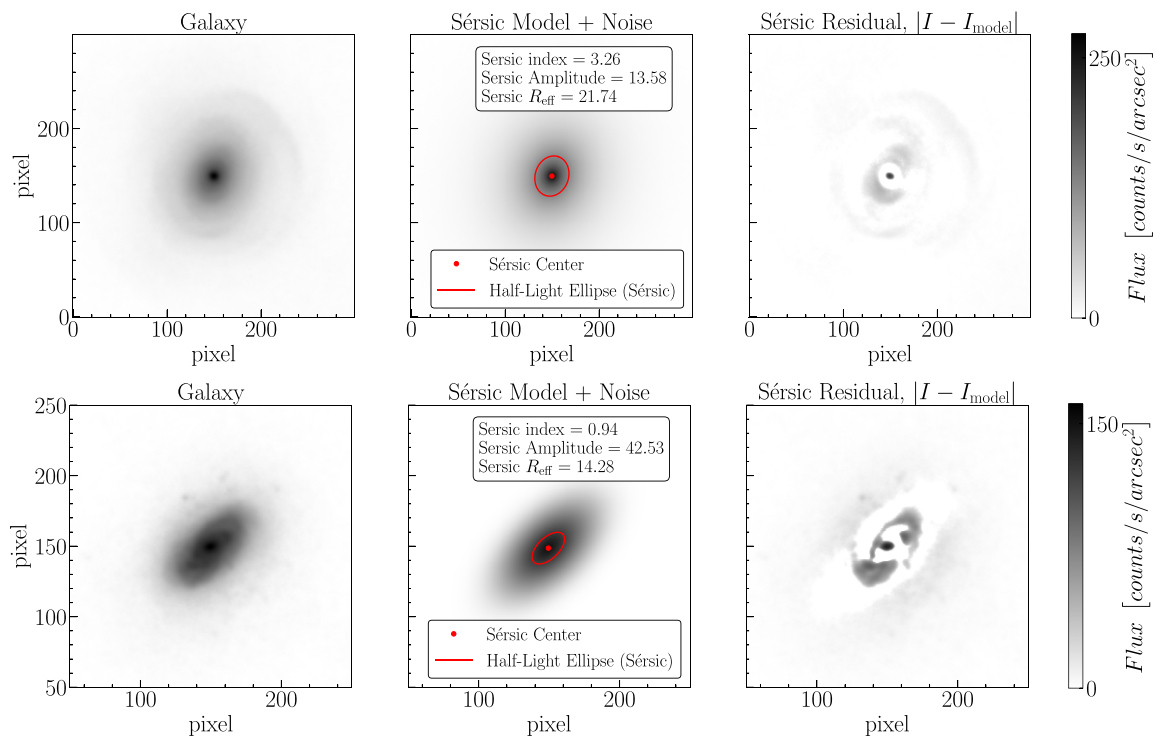

**Figure 9.** The *r*-band image, the Sérsic 2D best-fitting model, and the residuals, for the usual two example galaxies 96-2 (upper panels) and 96-3 (*bottom panel*). The best values for the Sérsic index, amplitude, and $R_{\rm eff}$ for the best-fitting profiles are given in the *central panels*. The amplitude is in counts arcsec$^{-2}$, while the $R_{\rm eff}$ is in pixel (these values can also be found in Table 1). Central panels also show the centre of the Sérsic 2D profile (red dot), and the half-light ellipse (red solid line).

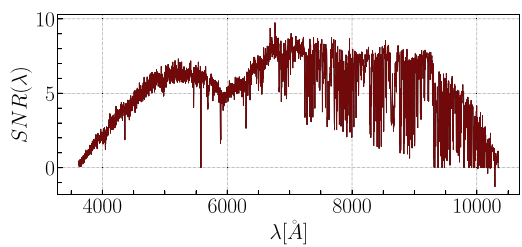

**Figure 10.** Averaged S/N as a function of the wavelengths across 100 randomly selected MaNGA galaxies, built from 100 MaNGA *LOGCUBE* considering all spaxels at 1.5$R_{\rm eff}$ and applying the *DRP3PIXMASK* mask in order to reject bad spaxels (Law et al. 2016).

The convolution of the data cubes for the reconstructed ePSF is the final step in the generation of a mock MaNGA galaxy.

### 3.5 Analysis of mock MaNGA data cubes

At this point, our mock data cubes have characteristics as close as possible to the MaNGA ones. We therefore proceed to analyse the mock data as if they were real data by following the MaNGA DAP (Belfiore et al. 2019; Westfall et al. 2019).

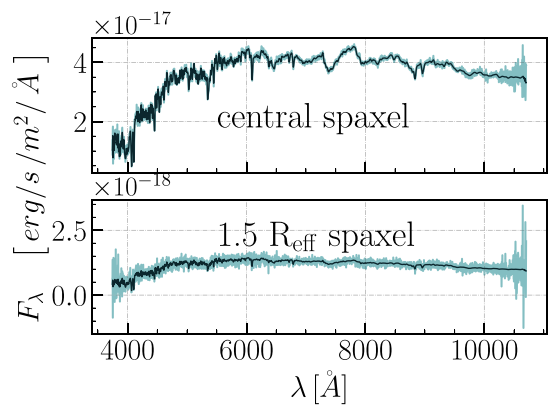

**Figure 11.** Effect of noise on mock spectra in two different spaxels for the example galaxy 96-3. The *upper panel* shows the spectrum in the central spaxel before (black line) and after adding noise as from equation (4) (light-blue line). The *bottom panel* shows the same for a spaxel at $\sim$1.5$R_{\rm eff}$.

#### 3.5.1 Voronoi binning

To extract unbiased measurements of stellar kinematics and stellar population properties, a good S/N is required. For this reason, the MaNGA DAP bins spectra by averaging neighbouring spaxels to meet a given S/N threshold. To this purpose, the adaptive spatial-binning scheme implemented by the Voronoi algorithm of Cappellari & Copin (2003) is exploited. As in the MaNGA DAP, we apply the Voronoi binning to achieve a minimum S/N of about 10 in the *g*-band image. To run the Voronoi algorithm as in the DAP, we construct the *g*-band noise and image by convolving by the SDSS *g*-band filter response function (Gunn et al. 1998) our data cube. We then follow the same further steps, by masking every pixel with $S/N_g < 1$ and computing the covariance between adjacent spaxels with the same approximations used in Westfall et al. (2019). The computation of the covariance between adjacent spaxels is important since without it fewer spaxels are binned together to reach the threshold of $S/N_g = 10$. As stressed in Westfall et al. (2019), even after applying the Voronoi binning, some of the tassels might not reach the given threshold. This is an effect of the algorithm: Not every S/N function can partition the FoV into tassels with equal S/N. Therefore, spectra with an S/N below 10 have to be expected.

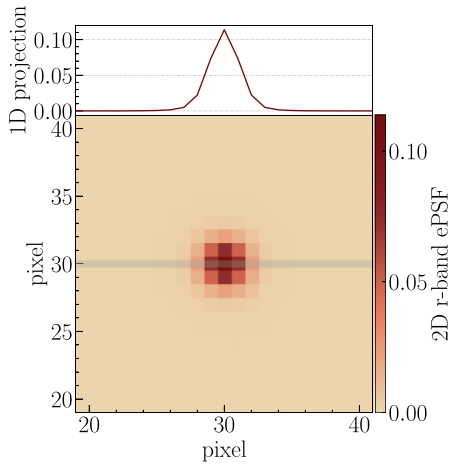

**Figure 12.** The *r*-band mock ePSF used in this work, as the average of 100 randomly selected *r*-band ePSFs in MaNGA *LOGCUBE*, from the MaNGA DRP outputs. The reconstructed ePSF is normalized so that it integrates to 1. On top, the projection of this 2D kernel along the slice highlighted with a grey band, as an example of its 1D profile.

Fig. 13 shows the *g*-band S/N maps before and after the Voronoi binning algorithm is applied (left- and right-hand panels, respectively), for galaxy 96-2 (top panels) and 96-3 (bottom panels). As expected, the S/N maps show higher values in the centre of the galaxies, as already discussed in Section 3.4 and demonstrated in Fig. 11. Therefore, the Voronoi binning has higher effects in the outskirts of the maps, where we can notice how some neighbouring pixels are averaged together (right-hand panels). Masking every pixel with $S/N_g < 10$ does not produce effects on these selected galaxies, since their S/N is always higher; therefore, no spaxels are rejected in the following analysis. The Voronoi tassellation computed for the *g*-band image is then applied to the entire data cube. We have 1074 Voronoi tassels for galaxy 96-2 and 1218 for galaxy 96-3.

### 3.5.2 Mock kinematics with PPXF

As in the MaNGA DAP, after applying the Voronoi binning, we run PPXF (Cappellari 2017) to analyse galaxy kinematics and model emission lines. This algorithm implements the Penalized Pixel-Fitting method (PPXF) to extract stellar and gas kinematics from spectra, using a maximum penalized likelihood approach. PPXF is a full-spectrum fitting algorithm: It assumes a galaxy spectrum is composed of a mixture of template spectra, convolved with the LOS velocity distribution (LOSVD) function of the stellar/gas kinematics.

In the MaNGA DAP, PPXF is exploited to study the stellar kinematics, the gas kinematics and the gas emission lines (Belfiore et al. 2019; Westfall et al. 2019). With the first two iterations, only the stellar kinematics is analysed. To run PPXF for measuring stellar kinematics, we use the MILES-HC libraries as templates, as in the MaNGA DAP, and we follow the same steps for the template preparation. Following Westfall et al. (2019), each fit iteration of PPXF uses an additive eighth-order Legendre polynomial and a Gaussian LOSVD. Both for the stellar continuum and the emission lines, just the first two moments are investigated. We also apply the same masking for pixel with $S/N_g < 1$. This first PPXF run must be performed over the stellar continuum only, therefore the emission lines are masked. Following the DAP, we mask the $\pm 750$ km s$^{-1}$ region around the 21 emission lines reported in table 3 in Westfall et al. (2019). To analyse the kinematics from the stellar continuum, PPXF is run twice: The first iteration fits the masked average of all the spectra in the data cube to select the subset of templates allocated non-zero weight; the second time, all tassels are fitted using just this subset of templates allocated non-zero weight. Excluding the subset of templates characterized by non-zero weight during the first iteration, speeds up the second iteration, but also limits the effect of noise-driven inclusion of templates when analysing spectra with lower $S/N_g$. In particular, we match the template-library resolution to the MaNGA data by convolving each spectrum with a wavelength-dependent Gaussian kernel (as explained in Appendix A in Westfall et al. 2019). Then, with a second full-spectrum fit, emission line profiles are modelled. This time, PPXF is exploited to fit at the same time all emission-line features and the stellar continuum. We prepare the templates following the MaNGA DAP, so considering MILES-HC for the stellar continuum as before, and also constructing the templates for the emission lines (as in section 9.1 in Westfall et al. 2019). We follow the steps in the DAP to create the equivalent of *VOR10-GAU-MILESHC* data for our galaxies: therefore, also for the emission-line fitting, we run PPXF over the Voronoi tasselled data cube.

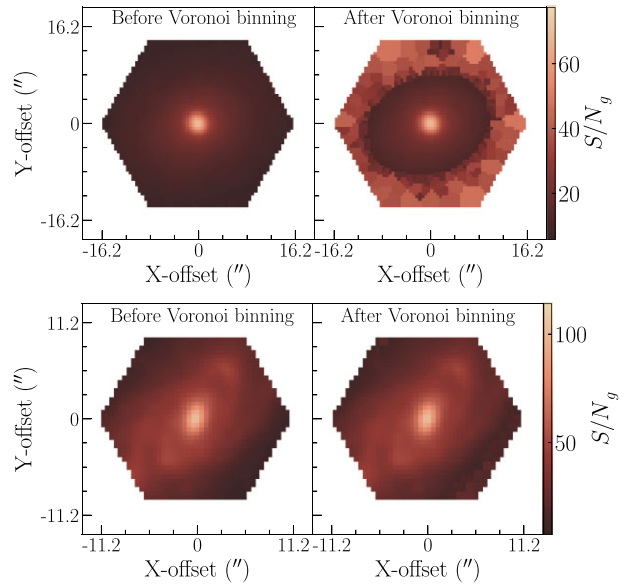

**Figure 13.** The *g*-band S/N map, before (*left panels*) and after (*right panels*) applying the Voronoi binning with target $S/N_g = 10$, for two mock MaNGA data cubes corresponding to the two example galaxies 96-2 (upper panels) and 96-3 (bottom panels), observed with the MaNGA FoV = 32.5 and 22.5 arcsec, respectively.

Thanks to PPXF, we are able to recover kinematic maps and the gas emission lines for our mock MaNGA galaxies as for MaNGA real galaxies. As a sanity check of our method, we compare our PPXF-derived kinematics with the intrinsic kinematics from TNG50.

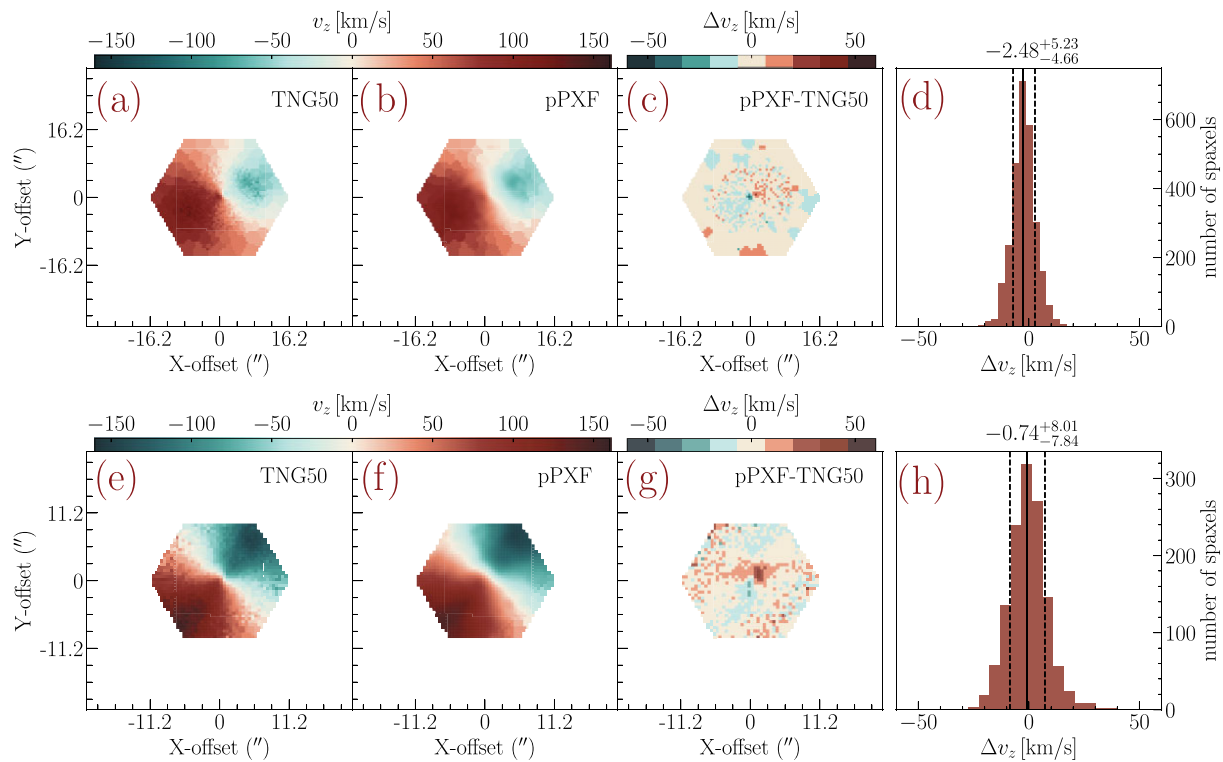

**Figure 14.** Intrinsic and PPXF-recovered stellar velocity maps, residual maps, and residual distributions for two Voronoi-binned mock MaNGA data cubes (see Sections 3.4–3.5.1) for galaxy 96-2 (upper row) and 96-3 (bottom row), observed with the MaNGA FoV = 32.5 and 22.5 arcsec, respectively. Panels **(a)** and **(e)** show maps of the stellar velocity along the LOS computed from TNG50, while panels **(b)** and **(f)** show the same as recovered with PPXF. Panels **(c)** and **(g)** show the maps of velocity residuals $\Delta v_z = v_{z,\mathrm{ppxf}} - v_{z,\mathrm{TNG50}}$, and panels **(d)** and **(h)** show the residual distributions and the 0.16, 0.5, and 0.84 quantiles. Over all the tassels, residuals are consistent with 0 at the 68 per cent confidence intervals, therefore stellar velocities recovered by running PPXF over the mock data cubes do not exhibit any systematic bias.

To compute the peculiar stellar velocity maps from TNG50, we apply the Voronoi grid over the simulated stellar particle distribution and measure the mass-weighted velocity along the *z*-axis in each tassel as

$$v_{z,\mathrm{tassel}} = \frac{\sum_{i=1}^{N_*} M_{i,*} \times v_{z,i}}{\sum_{i=1}^{N_*} M_{i,*}}, \qquad (5)$$

where $v_{z,i}$ is the velocity along the *z*-axis for the *i*-stellar particle of the tassel with stellar mass $M_{i,*}$, and $N_*$ is the number of stellar particles in that tassel.

From equation (5), the mass-weighted stellar velocity dispersion map is computed as the standard deviation of the peculiar stellar velocity in each tassel.

Fig. 14 shows the results of this test for galaxies 96-2, top rows, and 96-3, bottom rows. Panels **(a)** and **(b)**, and **(e)** and **(f)** show the peculiar stellar velocity along the LOS as computed from TNG50 (from equation (5)), and as recovered by running PPXF over the Voronoi-binned mock data cube. Maps are colour-coded according to the velocity values. Panels **(c)** and **(g)** show the maps of the residuals $\Delta v_z = v_{z,\mathrm{PPXF}} - v_{z,\mathrm{TNG50}}$, while panels **(d)** and **(h)** report the distributions of these residuals $\Delta v_z$ and the 0.16, 0.5, and 0.84 quantiles. We see that residuals are consistent with 0 at the 68 per cent confidence intervals, with no systematic bias, for both simulated galaxies. In summary, by running PPXF over the stellar continuum in each tassel following the same approach as in the MaNGA DAP, we are able to recover the stellar peculiar velocity along the LOS without introducing any bias.

Fig. 15 is the same as Fig. 14 for the mass-weighted stellar velocity dispersion and residuals as $\Delta \sigma_{v_z} = \sigma_{v_z,\mathrm{pPXF}} - \sigma_{v_z,\mathrm{TNG50}}$. Also, in this case, the stellar velocity dispersion along the LOS is recovered without systematic bias by running PPXF on the mock data cubes. Somewhat higher residuals in some of the tassels is not surprising, both for the stellar velocity and the stellar velocity dispersion, since we are applying the observational effects such as the convolution with the ePSF (see Section 3.4).

At the beginning of this section we explained that we also run PPXF to fit emission lines. We are therefore able to reconstruct, for example, the widely used H $\alpha$ equivalent width (EW) maps, as done for real MaNGA galaxies (e.g. Kauffmann 2021; Stark et al. 2021). Fig. 16 illustrates the H $\alpha$ EW map (left-hand panel) for galaxy 96-3, which we show alongside a popular stellar population indicator, the D-4000 using its narrow definition $D_n4000$ (Balogh et al. 1999; Kauffmann et al. 2003). The latter is computed integrating the data cube in the range 4000–4100 and 3850–3950 Å (right-hand panel). It is possible to notice that the H $\alpha$ EW drops at the centre of the galaxy, while it increases following the spiral structure. On the other end, the $D_n4000$ map shows an opposite trends to the H $\alpha$ EW map, as expected as this spectral index increases with age and metallicity. Here we show that we are producing realistic spectral indices such H $\alpha$ and $D_n4000$ for our mock MaNGA galaxies. Moreover, we verified that the locations with higher values of H $\alpha$ EW match the positions in the simulated galaxy where SFR > 0, as expected, showing the effectiveness of the MIII spectral models. There is a clear correlation between H $\alpha$ EW and SFR. More indices will be considered in follow-up papers.

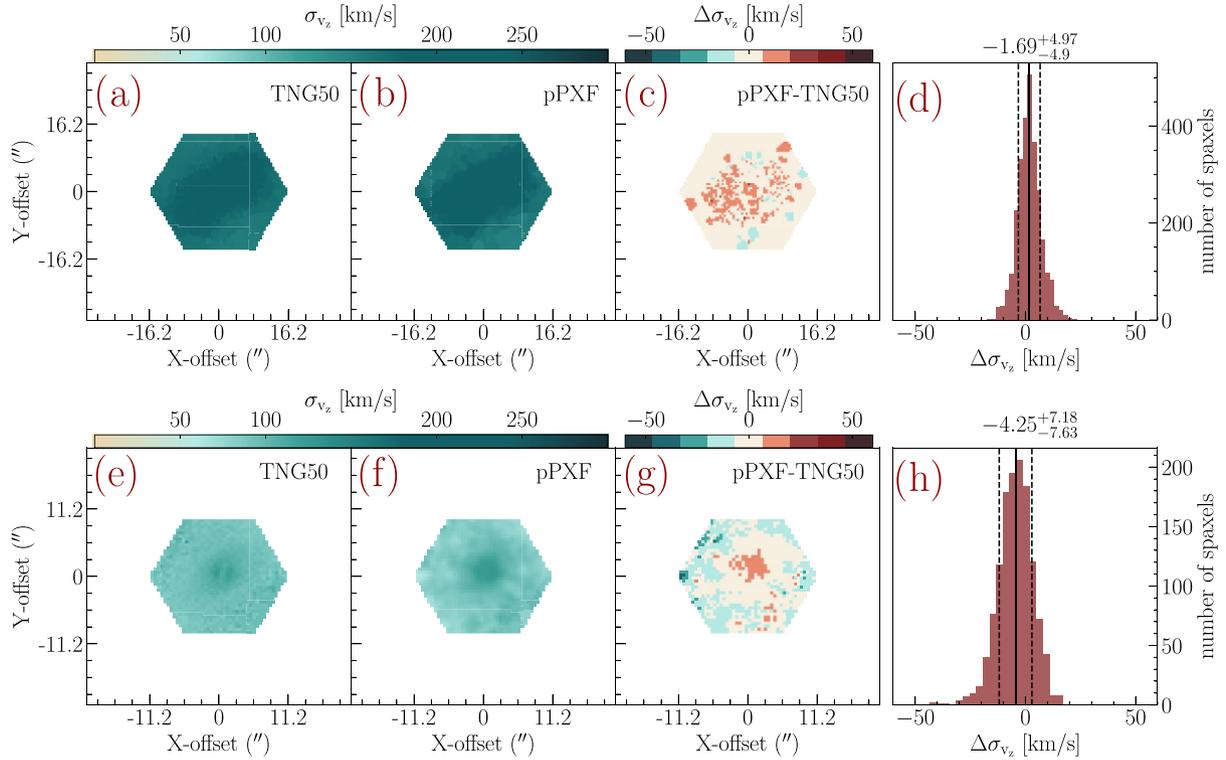

**Figure 15.** Same as Fig. 14 for the velocity dispersion maps and residuals.

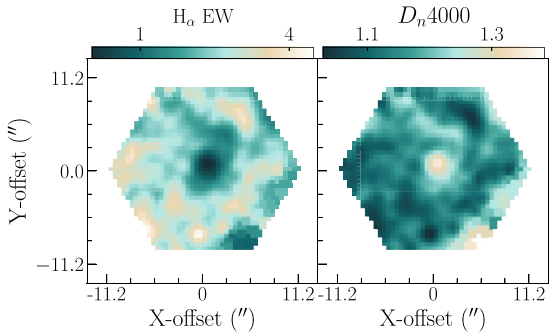

**Figure 16.** H$\alpha$ and $D_{n4000}$ EW maps for a mock MaNGA data cube with the MaNGA FoV = 22.5 arcsec for the example galaxy 96-3. The H$\alpha$ EW map is reconstructed from the emission-line best fit by pPXF, while the $D_{n4000}$ map is computed integrating the mock MaNGA data cube in the range 4000–4100 and 3850–3950 Å. As expected, they show opposite trends, since a stronger $D_{n4000}$ indicates older and more metal-rich stellar populations.

These comparisons are meant to test our ability to imprint the kinematics and the star-forming gas properties in the spectra. Therefore, we directly compare the observationally derived kinematics with the true kinematics of the galaxy, having applied the same FoV and tessellation. Other approaches have been adopted, as for example in Ibarra-Medel et al. (2018), where the scope was to test the ability to recover stellar properties from spectral fitting with PIPE3D (Sánchez et al. 2016); hence, observational effects such as dust extinction and the PSF effects were added to the truth tables to reconstruct the Pipe3D bias.

We now proceed to perform full spectral fitting of population models to our mock data cubes. The Voronoi tassels, stellar kinematics, and emission-line best fits will be used when running FIREFLY, following the same approach as in Neumann et al. (2021, 2022) for the analysis of real MaNGA galaxies with FIREFLY (Wilkinson et al. 2017).

# 4 RESULTS: ANALYSIS OF MANGA MOCK GALAXIES

In this section, we describe how we obtain the stellar population properties of our mock MaNGA data cubes and how these compare with the intrinsic properties from TNG50. These comparisons are not meant to test the FIREFLY ability to recover the stellar population properties, but our ability to imprint them into our spectra, as already discussed in Section 3.5.2. Therefore, when comparing the recovered properties to the intrinsic ones, we only apply the same FoV and tessellation to the truth tables.

## 4.1 Full spectral fitting with FIREFLY

We analyse the spectra of the mock data cubes with FIREFLY (Wilkinson et al. 2017), a chi-squared minimization spectral fitting code. FIREFLY obtains stellar population properties such as age, metallicity, stellar and remnant mass, and the star formation history (SFH), for a given input spectrum, by comparing it with combinations of single-burst stellar population models (SSPs) until convergence is achieved.

Specifically for IFU data as in this work we follow the same procedure that is adopted to perform spectral fitting for the real MaNGA data cubes as in Neumann et al. (2021). In summary, FIREFLY reads the Voronoi-binned spectra, subtracts the emission line fluxes, downgrades the resolution of the models to the combined instrumental resolution and velocity dispersion of the spectra, and shifts the spectra to rest-frame wavelengths using redshift and stellar velocity from our mock data cubes. As fitting templates we adopt the MaStar stellar population models (as in Neumann et al. 2021, 2022).

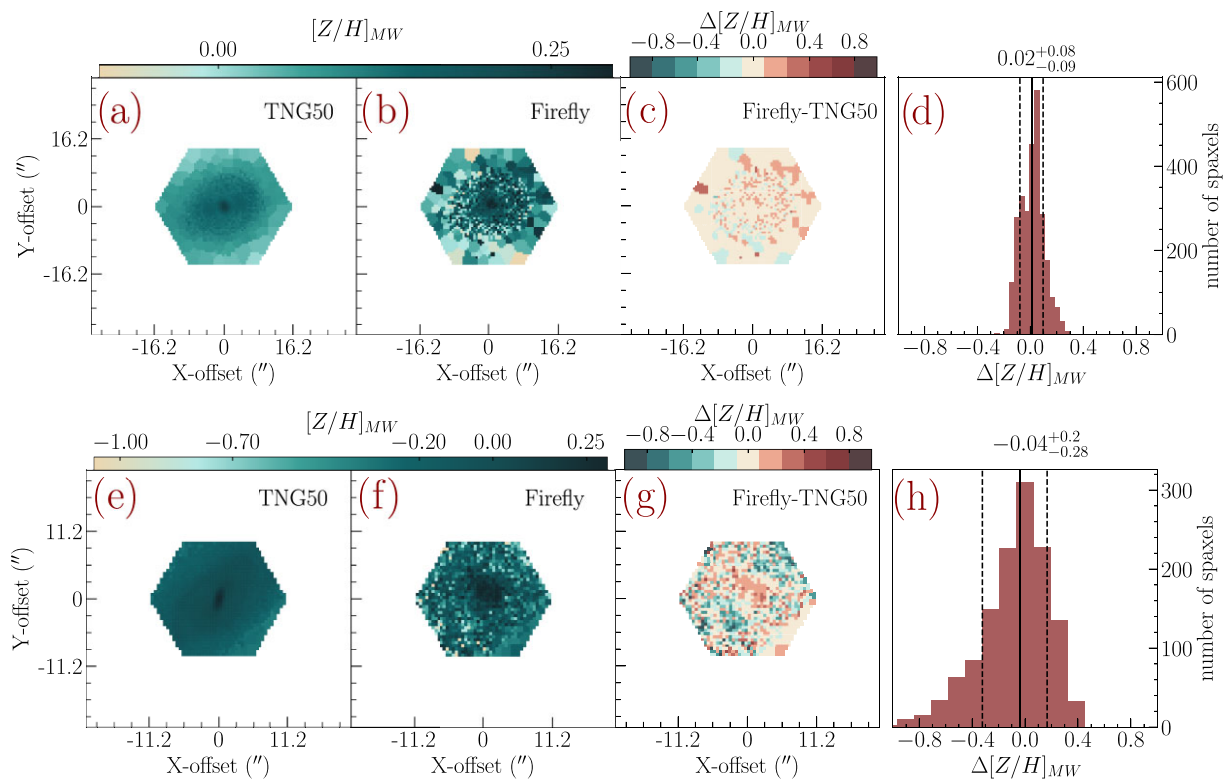

**Figure 17.** Metallicity maps from TNG50 (panels **a** and **e**) and FIREFLY (panels **d** and **f**), maps of residuals (panels **c** and **g**), colour-coded as $\Delta[Z/H]_{MW} = [Z/H]_{MW,PPXF} - [Z/H]_{MW,TNG50}$ and residual distributions (panels **d** and **h**) with 0.16, 0.5, and 0.84 quantiles, for Voronoi-binned mock MaNGA data cubes for TNG50 galaxies 96-2 (upper panels) and 96-3 (bottom panels) with MaNGA FoV = 32.5 and 22.5 arcsec, respectively. Over all tassels, residuals are consistent with 0 at the 68 per cent confidence intervals over all tassels, meaning that metallicity values are recovered without systematic bias by running FIREFLY over the mock MaNGA data cubes.

*4.1.1 Mass-weighted metallicity and age maps*

With FIREFLY, we are able to obtain several interesting output to perform galaxy evolution studies, for example, mass-weighted metallicity and age maps. In order to obtain the intrinsic mass-weighted metallicty $[Z/H]_{MW}$ and age $\log(Age)_{MW}$ maps for the simulated galaxies from TNG50, we apply the Voronoi grid over the simulated stellar particles and obtain the mass-weighted properties in each tassel as

$$\theta_{\text{WM,tassel}} = \frac{\sum_{i=1}^{N_*} M_{*,i} \times \theta_i}{\sum_{i=1}^{N_*} M_{*,i}}, \quad (6)$$

where $\theta_{\text{MW,tassel}}$ represents $Z/H_{MW}$ or $Age_{MW}$ in the selected tassel, $\theta_i$ represents the age or metallicity value for the *i*-particle of the considered tassel with stellar mass $M_{*,i}$, and $N_*$ is the number of stellar particles in the selected tassel.

Fig. 17 shows the mass-weighted metallicity $[Z/H]_{MW}$ in each Voronoi tassel as measured by FIREFLY (panels **a** and **e** for galaxy 96-2 and 96-3, respectively), and the intrinsic values from TNG50 for the same galaxies (panels **b** and **f**) computed with equation (6). Maps are colour-coded according to the $[Z/H]_{MW}$ values. Panels **(c)** and **(g)** show the maps of the differences as 'recovered - intrinsic', colour-coded according to the residual $\Delta[Z/H]_{MW} = [Z/H]_{MW,Firefly} - [Z/H]_{MW,TNG50}$. Panels **(d)** and **(h)** report the residual $\Delta[Z/H]_{MW}$ distributions and their 0.16, 0.5, and 0.84 quartiles.

While residuals are very small and well within the error for the 'passive' galaxy 96-2 (panel **c**), we find larger residual in some tassels for the star-forming galaxy 96-3 (panel **g**). This is expected, as it is generally more difficult to recover the properties of galaxies with ongoing star formation (e.g. Pforr, Maraston & Tonini 2012; Wilkinson et al. 2017) due to age-dust-metallicity degeneracies – galaxy 96-3 does contain dust – and the overshining effect from the youngest stellar generations (Maraston & Strömbäck 2011). Overall however, residuals are consistent with 0 at the 68 per cent confidence interval for both galaxies (panels **d** and **h**).

Fig. 18 is the analogue to Fig. 17 for the mass-weighted age maps and residuals. Also in this case, galaxy 96-3 exhibits larger residuals, as expected for the same reasons discussed above. Also in the case of age, over all the tassels, residuals are consistent with 0 at the 68 per cent confidence interval.

These results are encouraging and suggest that our method does not carry particular biases into the analysis. Some non-null residuals are to be expected both because of well-known and complicated spectral degeneracies and because we add observational effects to the model data cubes (see Section 3.4) that degrade the intrinsic signal. Clearly, with just two example galaxies we cannot assess the size of these effects. This will be subject of future work on the analysis of the full TNG50 catalogue.

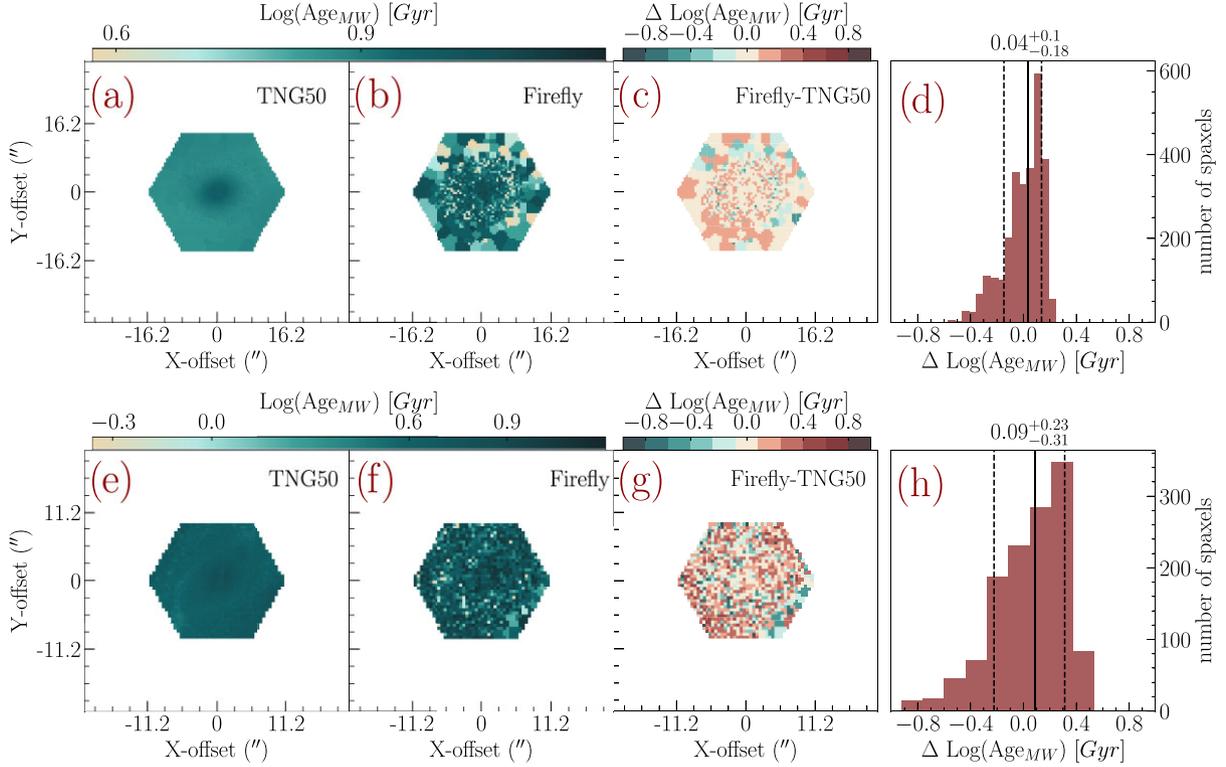

**Figure 18.** As in Fig. 17 for the mass-weighted age maps and residuals.

### 4.1.2 Mass-weighted metallicity and age gradients

Another crucial property signposting the evolution of galaxies are radial gradients in stellar populations, one of the main scopes for performing IFS. Following Neumann et al. (2022), we compute the radial gradients of age and metallicity. A gradient is calculated by first running a median along 10 equally sized radial bins and subsequently defining the gradients as

$$\nabla \theta =د\theta / dR, \quad (7)$$

where $R$ is the radius in units of effective radius $R_{eff}$ and $\theta$ is the binned median stellar population property (either $[Z/H]_{MW}$ or $\log(Age)_{MW}$). The radius $R$ is computed as the on-sky distance of each Voronoi cell from the centre of the galaxy. The gradient is measured using the least-squares linear regression. Errors on the gradients, $\sigma(\Delta\theta)$, are calculated using a Monte Carlo bootstrap resampling method, namely we iterate the boostrap resampling 1000 times building a distribution of gradient values, and use the standard deviation of this distribution as the 1$\sigma$ error on the gradient.

We also evaluate the intrinsic gradients from TNG50 (applying equations 6–7) and compare recovered (from FIREFLY) and intrinsic by evaluating the normalized residuals as

$$\frac{\nabla\theta_{TNG50} - \nabla\theta_{Firefly}}{\sqrt{\sigma(\nabla\theta_{TNG50})^2 + \sigma(\nabla\theta_{Firefly})^2}}. \quad (8)$$

Results are shown in Fig. 19, for galaxy 96-2 (top row) and 96-3 (bottom row). The left-hand panels show the mass-weighted metallicity gradients obtained from full spectral fitting (dashed black lines) and intrinsic to the TNG50 simulations (solid yellow), as obtained using equation (7). The green dots are the mean mass-weighted metallicity values estimated by FIREFLY in each tassel with their 1$\sigma$ error as a function of radius given as $R/R_{eff}$. Diamonds show the binned median mass-weighted metallicity values from FIREFLY, with errorbars from the boostrap method. The normalized residuals for both galaxies, computed as in equation (8), are <1; therefore, recovered and intrinsic gradients are statistically consistent within 1$\sigma$.

The right-hand panels in Fig. 19 show the mass-weighted age gradients, with identical meaning of symbols as in the left-hand panels. Also in case of the age, recovered and intrinsic gradients are statistically consistent within 1$\sigma$ (normalized residuals are <1).

From these results (Figs 17–19), we can see how galaxy 96-2 is characterized by metal-rich stellar populations in the centre. Galaxy 96-3 has a steeper metallicity gradient, with the outskirts characterized by very metal poor stellar populations. Moreover, both galaxies exhibit flat age gradients. In general, galaxy 96-3 contains younger stellar populations overall compared to galaxy 96-2.

### 4.1.3 Star formation histories

The SFH is one of the most challenging properties to be recovered from data and yet a crucial quantity to constrain galaxy formation and evolution processes. The FIREFLY code allows us to resolve the SFH non-parametrically in each tassel, which we can then compare to the intrinsic one from TNG50. Fig. 20 (left-hand panels) shows the SFH as SSP mass-weights versus look-back time, as recovered by FIREFLY (green) and as intrinsic to the TNG50 simulations, for TNG50 galaxy 96-2 (upper panel) and 96-3 (lower panel), for all stellar particles within the FoV. To ease the comparison, the right-hand panels plot the differences between SSP mass-weights as 'recovered - intrinsic' (i.e. 'FIREFLY - TNG50').

Although for both galaxies, the full spectral fitting recovers a fraction of old (14 Gyr) and young (3 Gyr) population components that are absent in the TNG50 output, the overall shape of the SFH is well recovered, suggesting an exponential decay for the passive galaxy and a nearly constant star formation for the star-forming galaxy, resembling sufficiently well the intrinsic histories.

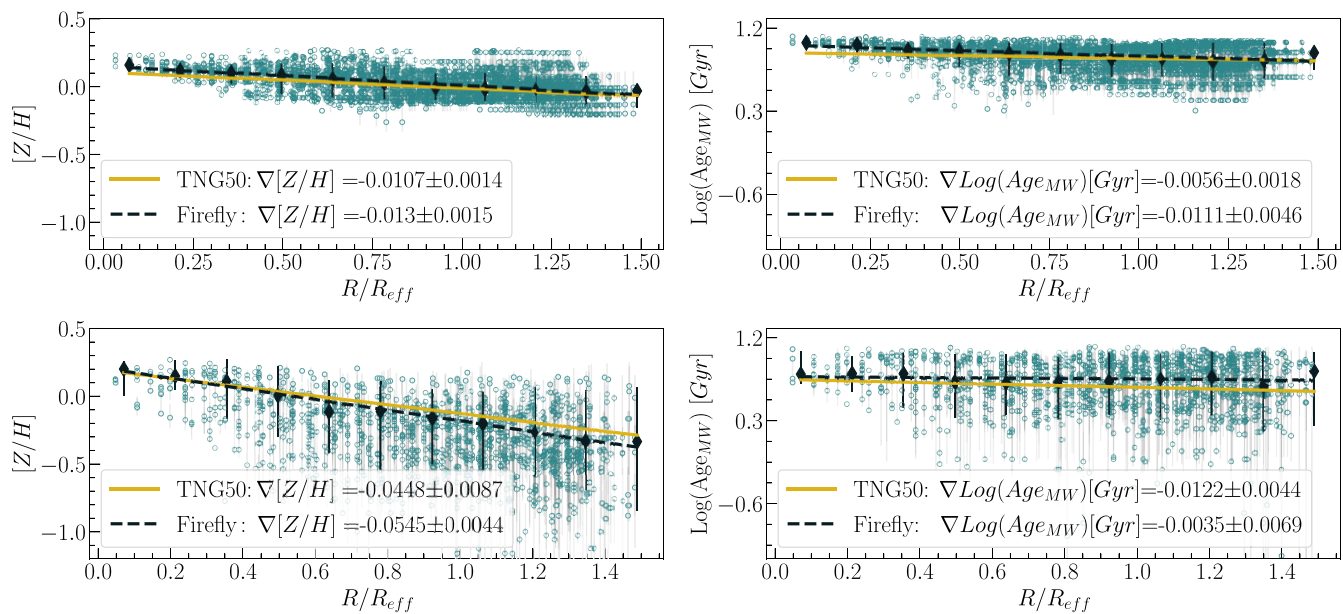

**Figure 19.** Gradients in metallicity (left-hand panels) and age (right-hand panels), from TNG50 as solid yellow lines and FIREFLY as dashed black lines, for two Voronoi-binned mock MaNGA data cubes for TNG50 galaxies 96-2 (upper panels) and 96-3 (bottom panels). Green circles with grey error bars are the values output of FIREFLY) and black diamonds are the binned median values, with errorbars from boostrapping. Normalized residuals, as from equation (8), for both $\theta = [Z/H]_{MW}$ and $\theta = \log(Age)_{MW}$, are $<1$; therefore, the intrinsic gradients from TNG50 and those recovered from FIREFLY are statistically consistent within $1\sigma$.

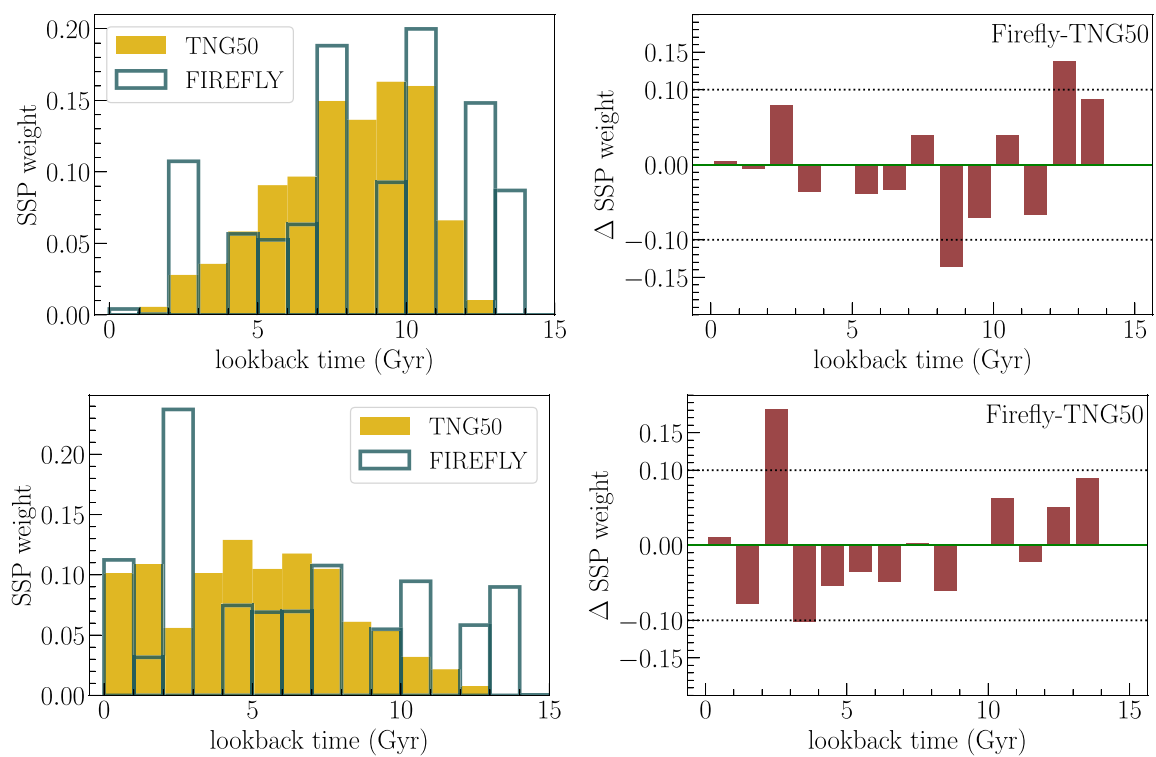

**Figure 20.** SFHs resolved by FIREFLY (green histograms) compared to those intrinsic to the TNG50 simulations (yellow histograms) for the two example galaxies 96-2 (upper panels) and 96-3 (bottom panels). SFHs are represented as SSP mass-weights as a function of the lookback time. The right-hand panels display the differences in SFH as 'FIREFLY - TNG50' for an easier comparison.

### 4.1.4 Mass and E(B − V) maps

As a last output, we show the colour excess $E(B-V)$ and stellar mass maps as recovered through full spectral fitting with FIREFLY in

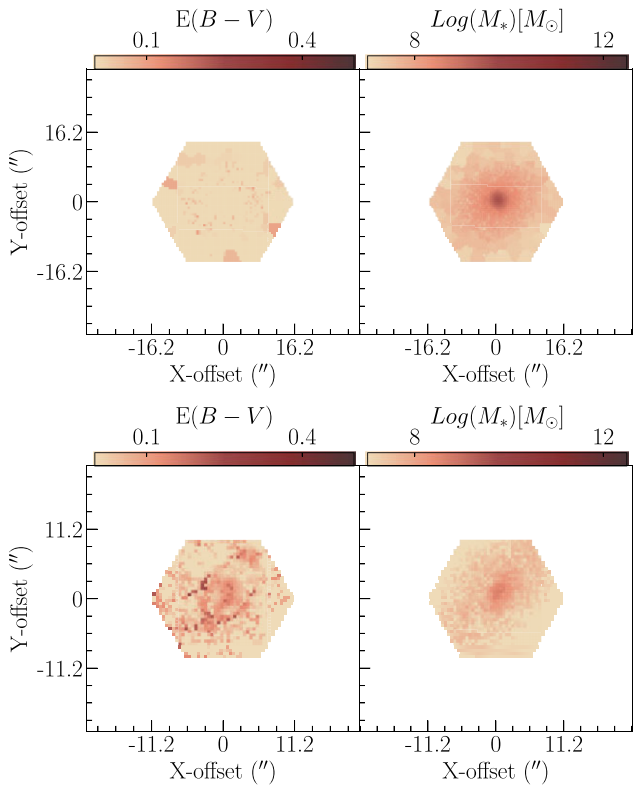

**Figure 21.** $E(B - V)$ and stellar mass maps obtained from FIREFLY for TNG50 galaxy 96-2 (upper panels) and 96-3 (bottom panels), observed with MaNGA FoV = 32.5 and 22.5 arcsec, respectively.

Fig. 21 (for galaxy 96-2 in the top panels and 96-3 in the bottom panels). It is reassuring to see that for galaxy 96-3, we recover high values for the $E(B - V)$, which are distributed along a structure resembling a disc morphology. This is consistent with our modelling of dust effects being proportional to the star-forming gas component using SKIRT outputs (as discussed in Section 3.2.2). For galaxy 96-2, instead, which has no ongoing star formation, reddening is essentially zero.

It can also be appreciated that galaxy 96-2 has a higher stellar mass than galaxy 96-3, consistent with the TNG50 values (Table 1). Moreover, while the stellar mass in galaxy 96-2 is homogeneously distributed, galaxy 96-3 displays an elongated distribution of stellar mass, as expected from its morphological shape.

## 5 SUMMARY AND CONCLUSIONS

In this paper, we have outlined and demonstrated our method to post-process *state-of-the-art* magneto-hydrodynamical galaxy simulations such as Illustris and IllustrisTNG to obtain mock MaNGA data cubes and to analyse them with the same techniques used in the MaNGA DAP (Westfall et al. 2019). For obtaining the spectral energy distributions of simulated galaxies, we adopt new stellar population models (Maraston et al. 2020) that are based on the MaNGA stellar library (MaStar, Yan et al. 2019), meaning based on the same instrument used to obtain the galaxy data. We further utilize the code SKIRT to include dust effects in the simulations. We finally perform full spectra fitting with FIREFLY (Wilkinson et al. 2017) to recover the stellar population properties of our mock data cubes.

To each stellar particle of a simulated galaxy, a model spectral energy distribution is assigned according to its mass, age, and metallicity (Section 3.2.1). For stellar particles younger than 4 Myr, we further include a modelling of the star-forming nebula using MIII models (Groves et al. 2008). For each spaxel, the Doppler effect, considering both the peculiar velocity along the LOS (the $z$-axis centred in the centre of mass of the simulated galaxies) and the galaxy redshift, is included. Spectra are then downgraded, according to the velocity dispersion along the LOS and the BOSS spectrograph resolution. Gas kinematics are included if appropriate. When dust is present, the attenuation curve in each spaxel is reconstructed using SKIRT (Section 3.2.2), and applied to the data cube.

We then simulated an IFU observation, positioning the instrument at the redshift of the galaxy in the simulation, along the $z$-axis, with an FOV of 150 × 150 arcsec$^2$ and pixel size of 0.5 arcsec as in real MaNGA observations, building a data cube with the same characteristic in spatial sampling as MaNGA. Thanks to our use of MaStar population models, the spectral resolution is equal to MaNGA observations as well. $r$-band SDSS images are then constructed in order to obtain the Sérsic morphology of our simulations using the code STATMORPH (Section 3.3). Once the effective radius for the simulated galaxy is known, one of the five fibre bundle configurations in MaNGA is applied to the galaxy. At that point, over the entire data cube, a reconstructed ePSF from MaNGA observations in each band is applied, as well as a noise based on the S/N($\lambda$) from MaNGA observations (Section 3.4).

Once all observational effects are included, we follow the procedure of the MaNGA Data Analysis Pipeline (DAP; Westfall et al. 2019), to re-bin the data cube, obtain the kinematics, and model emission lines (as explained in Section 3.5). The binning, kinematics, and emission lines were used to finally run the full spectral fitting code FIREFLY over the data cubes (Section 4), as done in other works for real MaNGA observations (e.g. Goddard et al. 2016; Goddard et al. 2017; Neumann et al. 2021, 2022).

The pipeline presented in this paper can be easily modified to generate mock galaxies from any other hydro-dynamical galaxy simulation. Depending on the type of observation, further effects may need to be implemented. The effect of sky-subtraction residuals on the modelled spectra in the pipeline, for instance, will be key when modelling objects from the upcoming SDSS-V Local Volume Mapper survey (Konidaris et al. 2020).

In this paper, we applied our procedure to two example galaxies from TNG50 with different physical properties (age, mass, morphology, and SFH). We then compared the recovered properties to the intrinsic ones from TNG50. This analysis informs us on both the robustness of our mock data cube creation as well as any bias introduced by the spectral fitting procedure, which will be important to take into account for when we will use the simulations to constrain galaxy formation and evolution.

We find a generally good agreement between recovered and intrinsic properties, over all considered properties, namely kinematics, age and metallicity distributions and gradients, with residuals over all tassels consistent with 0 at the 68 per cent confidence level without any systematic bias. We also obtain star formation histories closely resembling the truth values.

In the next paper of this series we will present an entire mock MaNGA galaxy catalogue with ∼10 000 IFU simulations from TNG50. With this tool, we will carry out a systematic comparison between mock and real MaNGA galaxies in order to shed light on the physical processes of galaxy formation and evolution and to understand and constrain the astrophysics of modern cosmological simulations.

**ACKNOWLEDGEMENTS**

LN was supported by an STFC studentship. STFC is acknowledged for support through the Consolidated Grant Cosmology and Astrophysics at Portsmouth, ST/S000550/1. Numerical computations were done on the Sciama High Performance Compute (HPC) cluster that is supported by the ICG, SEPnet, and the University of Portsmouth. Funding for the Sloan Digital Sky Survey IV has been provided by the Alfred P. Sloan Foundation, the US Department of Energy Office of Science, and the Participating Institutions. SDSS-IV acknowledges support and resources from the Center for High Performance Computing at the University of Utah. The SDSS website is www.sdss.org. SDSS-IV is managed by the Astrophysical Research Consortium for the Participating Institutions of the SDSS Collaboration including the Brazilian Participation Group, the Carnegie Institution for Science, Carnegie Mellon University, Center for Astrophysics | Harvard & Smithsonian, the Chilean Participation Group, the French Participation Group, Instituto de Astrofísica de Canarias, The Johns Hopkins University, Kavli Institute for the Physics and Mathematics of the Universe (IPMU)/University of Tokyo, the Korean Participation Group, Lawrence Berkeley National Laboratory, Leibniz Institut für Astrophysik Potsdam (AIP), Max-Planck-Institut für Astronomie (MPIA Heidelberg), Max-Planck-Institut für Astrophysik (MPA Garching), Max-Planck-Institut für Extraterrestrische Physik (MPE), National Astronomical Observatories of China, New Mexico State University, New York University, University of Notre Dame, Observatário Nacional/MCTI, The Ohio State University, Pennsylvania State University, Shanghai Astronomical Observatory, United Kingdom Participation Group, Universidad Nacional Autónoma de México, University of Arizona, University of Colorado Boulder, University of Oxford, University of Portsmouth, University of Utah, University of Virginia, University of Washington, University of Wisconsin, Vanderbilt University, and Yale University. The primary TNG simulations were realized with compute time granted by the Gauss Centre for Supercomputing (GCS): TNG50 under GCS Large-Scale Project GCS-DWAR (2016; PIs Nelson/Pillepich), and TNG100 and TNG300 under GCS-ILLU (2014; PI Springel) on the GCS share of the supercomputer Hazel Hen at the High Performance Computing Center Stuttgart (HLRS).

**DATA AVAILABILITY**

MaNGA data are part of SDSS-IV, publicly available at (Abdurro'uf et al. 2022). The FIREFLY code is available at https://www.icg.port.ac.uk/firefly/ (Wilkinson et al. 2017) and the MaStar population models at https://www.icg.port.ac.uk/mastar/ (Maraston et al. 2020). Illustris and IllustrisTNG data are publicly available at https://www.illustris-project.org/data/ (Nelson et al. 2021).

This paper has been typeset from a TEX/LATEX file prepared by the author.